\documentstyle[12pt]{article}
\def\theequation{\thesection.\arabic{equation}}
\let\ssection=\section
\def\section{\setcounter{equation}{0}\ssection}

\def\open#1{\mbox{{\bf{#1}}}} 

\font\Scbig=cmss10 scaled\magstep1
\font\Scscr=cmss8 scaled\magstep1
\font\Scscrscr=cmss8
\newfam\Scfam
\textfont\Scfam=\Scbig
\scriptfont\Scfam=\Scscr
\scriptscriptfont\Scfam=\Scscrscr
\def\Sc{\fam\Scfam}

\def\fr#1{\mbox{{\bf{#1}}}} 

\makeatletter

\newdimen\normalarrayskip              
\newdimen\minarrayskip                 
\normalarrayskip\baselineskip
\minarrayskip\jot
\newif\ifold             \oldtrue            \def\new{\oldfalse}
\def\arraymode{\ifold\relax\else\displaystyle\fi} 
\def\@arrayskip{\ifold\baselineskip\z@\lineskip\z@
     \else
     \baselineskip\minarrayskip\lineskip2\minarrayskip\fi}
\def\@arrayclassz{\ifcase \@lastchclass \@acolampacol \or
\@ampacol \or \or \or \@addamp \or
   \@acolampacol \or \@firstampfalse \@acol \fi
\edef\@preamble{\@preamble
  \ifcase \@chnum
     \hfil$\relax\arraymode\@sharp$\hfil
     \or $\relax\arraymode\@sharp$\hfil
     \or \hfil$\relax\arraymode\@sharp$\fi}}
\def\@array[#1]#2{\setbox\@arstrutbox=\hbox{\vrule
     height\arraystretch \ht\strutbox
     depth\arraystretch \dp\strutbox
     width\z@}\@mkpream{#2}\edef\@preamble{\halign \noexpand\@halignto
\bgroup \tabskip\z@ \@arstrut \@preamble \tabskip\z@ \cr}%
\let\@startpbox\@@startpbox \let\@endpbox\@@endpbox
  \if #1t\vtop \else \if#1b\vbox \else \vcenter \fi\fi
  \bgroup \let\par\relax
  \let\@sharp##\let\protect\relax
  \@arrayskip\@preamble}
\makeatother

\def\lvm{\leavevmode\hbox to\parindent{\hfill}}
\def\bar{\overline}
\def\req#1{(\ref{#1})}
\def\dd#1{{\partial\over\partial{#1}}}
\def\L{\left}
\def\R{\right}

\def\sumr{\sum_{r\geq1}}
\def\sums{\sum_{s\geq1}}
\def\sump{\sum_{p\geq-1}}
\def\sumji{\sum_{j\neq i}}
\def\sumki{\sum_{k\neq i}}

\def\BE{\begin{equation}}
\def\EE{\end{equation} \vskip 0.30\baselineskip}
\def\BA{\begin{array}}
\def\EA{\end{array}}

\def\noi{\noindent}

\def\frac#1#2{{\textstyle{{#1}\over{#2}}}}
\def\half{{1\over2}}

\def\Kr#1{\delta_{{#1},0}}
\def\ket#1{|{#1}\rangle}

\def\ccases#1#2{\L\{\!\new\BA{l}{#1}\\ {#2}\EA\R.}

\def\hn{\hat{n}}
\def\hL{\widehat{L}}

\def\tT{\widetilde{T}}

\def\cA{{\cal A}}
\def\cG{{\cal G}}
\def\cH{{\cal H}}
\def\cI{{\cal I}}
\def\cL{{\cal L}}
\def\cO{{\cal O}}
\def\cQ{{\cal Q}}
\def\cT{{\cal T}}
\def\cU{{\cal U}}
\def\cZ{{\cal Z}}

\def\oI{{\open I}}
\def\oL{{\open L}}
\def\oZ{{\open Z}}

\def\sI{{\Sc I}}
\def\sJ{{\Sc J}}
\def\sL{{\Sc L}}
\def\sQ{{\Sc Q}}
\def\sT{{\Sc T}}
\def\sU{{\Sc U}}
\def\sV#1{{{\Sc V}^{(#1)}}}

\def\ctop{{\Sc c}}
\def\htop{{\Sc h}}

\def\a{\alpha}
\def\b{\beta}
\def\g{\gamma}
\def\d{\partial}
\def\dd#1{{\partial\over\partial{#1}}}
\def\ddsc#1{{\partial^2\over\partial{#1}^2}}
\def\pr{^\prime}
\def\Rot{n_i\partial_j-n_j\partial_i}
\def\inv#1{{1\over{#1}}}

\def\cc{central charge}
\def\bc{background charge}
\def\tcc{topological central charge}
\def\ie{{i.e.}}

\def\de{decoupling equation}
\def\h{hierarchy}
\def\hs{hierarchies}
\def\cs{constraints}
\def\K{Kontsevich}
\def\M{Miwa}
\def\V{Virasoro}
\def\emt{energy-momentum tensor}

\begin{document}
\hfuzz=1pt

\let\ccirc=\circ
\def\circ{\,\raisebox{1pt}{{${\scriptstyle{\ccirc}}$}}\,}

\begin{flushright}
CERN-TH.6752/92
\end{flushright}
\vspace*{0.5mm}
\addtolength{\baselineskip}{.4\baselineskip}
\begin{center}
{\LARGE {\sc Singular Vectors and Topological Theories
from Virasoro Constraints via
the \K-\M\ Transform}}\end{center}
\vspace*{.5cm}

\addtolength{\baselineskip}{-.28571\baselineskip}
\begin{center}
{{\large {\bf B.~Gato-Rivera}}\\ {\small {\sl Instituto de
Matem\'aticas y F\'\i sica Fundamental, Serrano 123, Madrid 28006,
Spain}}\\{\mbox{}}\\
and\\ {\mbox{}}\\ {\large {\bf A.~M.~Semikhatov}}\\ {\small {\sl CERN,
CH-1211 Geneva 23, Switzerland}}\\ {\small and}\\ {\small {\sl
P.~N.~Lebedev Physics Institute}}\\ {\small {\sl Leninsky pr.\
53, Moscow 117924, Russia}}}\\ 
\end{center}
\vspace*{1cm}
\noindent
\begin{abstract}
{\small We use the \K-\M\ transform
to relate the different pictures describing matter coupled to
topological gravity in two dimensions: topological theories,
\V\ \cs\ on integrable \hs, and a DDK-type formalism.
With the help of
the \K-\M\ transform, we solve
the \V\ \cs\ on the KP \h\
in terms of minimal models
dressed with a (free) Liouville-like
scalar. The dressing prescription originates
in a topological (twisted $N\!=\!2$)
theory.
The Virasoro constraints are thus related
to essentially the $N\!=\!2$ null state decoupling equations.
The $N\!=\!2$ generators are constructed out of
matter, the `Liouville' scalar, and $c\!=\!-2$ ghosts.
By a `dual' construction involving the
reparametrization $c\!=\!-26$ ghosts, the
DDK dressing prescription is reproduced from the $N\!=\!2$ symmetry.
As a by-product we thus observe
that there are two ways to dress arbitrary
$d\!\leq\!1\bigcup d\!\geq\!25$
matter theory, which
allow its embedding into a topological
theory.
By the \K-\M\ transform, which introduces an infinite set
of `time' variables $t_r$, the equations ensuring the vanishing of
correlators that involve
BRST-exact primary states, factorize
through
the Virasoro generators expressed in terms of the $t_r$.
The \bc\ of these \V\ generators is
determined in terms of the \tcc\ $\ctop\!\neq\!3$ as
$\sQ=\sqrt{{3-\ctop\over3}}-2\sqrt{{3\over3-\ctop}}$\,.
}\end{abstract}
\thispagestyle{empty}

\vfill{\hbox to\hsize{CERN-TH.6752/92\hfill}
\hbox to\hsize{IMAFF-92/8\hfill}
\hbox to\hsize{hep-th@xxx/9212113\hfill}
\hbox to\hsize{December 1992\hfill}}

\newpage
\thispagestyle{empty}
\tableofcontents

\newpage

\setcounter{page}{1}

\addtolength{\baselineskip}{.4\baselineskip}
\parskip=0.2\baselineskip

\section{Introduction and discussion}
\subsection{Virasoro constraints as decoupling
equations}\lvm
An important result of matrix models
\cite{[BK],[DSh],[GM]} was the discovery of an `integrable' counterpart of
the topological gravity~+~matter theories in two dimensions,
in the form of {\it constrained\/} integrable
\hs\ \cite{[D],[FKN1],[DVV]}.
A central notion of the `integrable' formalism is the string
equation \cite{[D]},
or, which is essentially the same, the \V\ \cs\ \cite{[FKN1],[DVV]},
which must
encode the `dynamical' content of the theory and
therefore must somehow be related to continuum
field-theoretic formalism.

However, an immediate problem that one encounters
when attempting to construct
a direct correspondence
between the integrable and
the continuum formulations,
is how to interpret
in `intrinsic' CFT terms the
infinite collection of time parameters which label
the integrable evolutions (in the language of topological
theories the times appear as `external' parameters
that allow one to write down the generating
function for the amplitudes \cite{[K],[W2]}).
In terms of these times $t_{r\geq1}$, the most general
\V\ \cs\ on the tau function of the KP \h\
read
$\sL_p\tau=0$, $p\geq-1$, where

\BE\new\BA{rcl}\sL_{p>0} &=&\half\sum^{p-1}_{s=1}{\d^{2}\over\d t_{p-s}\d
t_s}+\sum_{s\geq 1}st_s {\d\over\d t_{p+s}}+\biggl(\sJ-\half\biggr)(p+1)
{\d\over\d t_p}\\ \sL_0&=&\sum_{s\geq 1}st_s {\d\over\d t_s}\\ {\Sc
L}_{-1}&=&\sum_{s\geq 1}(s+1)t_{s+1}{\d\over\d t_s} \EA\label{Lontau}\EE

\noi
with $\sJ$ being a free parameter\footnote{For $\sJ$ arbitrary,
the \cs\ are more general than those actually derived from
matrix models. However, our point of view is to promote to a
`first principle' the fact that appropriately
constrained integrable \hs\ provide an alternative
description of topological quantum gravity
\cite{[W-Dec89],[DW-Feb90],[VV-Apr90]}, which would
make it natural to
consider on an equal footing various
\cs, including those whose derivation
from a specific {\it matrix\/} model is
not known (yet).
This point of view has also
been advocated previously in
\cite{[S34],[HMG]}.}.

We will show that a conformal field theory
can be recovered as a {\sl solution\/} to these \V\ \cs.
To start with, recall that there does exist
a transformation,
known in the theory
of integrable systems as the Miwa transform
\cite{[Mi],[Sa]},
which expresses the (complexified)
KP times through
`complex coordinates' living on the spectral parameter worldsheet:

\BE t_r={1\over r}\sum_j n_jz^{-r}_j,\quad r\geq1\label{Miwatransform}\EE

\noi
where $\{z_j\}$ is a set (infinite in order that the $t_r$ be
independent) of points on the complex plane.
As far as the integrable systems were concerned,
the main interest in applications of \req{Miwatransform} was
concentrating around rewriting the integrable equations
(taken in the Hirota form \cite{[DDKM]})
as finite-difference equations in
the $n_j$. On the other hand, in a context much closer
to that of the present paper,
Kontsevich \cite{[K]} has used a
parametrization for the time parameters similar to \req{Miwatransform},
in which, however, the $z_j$, rather than
the $n_j$, were playing
the `active' role, while all the $n_j$ were set to a constant (see also
\cite{[MS]}--\cite{[IZ]}).

We will combine
these two points of view on the transformation \req{Miwatransform}
by regarding it
as a {\it set\/} of transformations from the complex coordinates
$z_j$ to the times $t_r$,
{\it parametrized\/} by the $n_j$, and call it the
\K-\M\ transform \cite{[S35]}.
The question now is how meaningful CFT data can be recovered
by applying \req{Miwatransform} to the KP \h\ constrained by
the operators \req{Lontau}.

A mathematical statement which we observe to
underlie the sought correspondence, is formulated in
Sect.~7 as a theorem on a class of differential operators.
However, we would prefer to adopt a more physical standpoint.
We are going to interpret the $z_j$ introduced via
eq.~\req{Miwatransform} as positions of
certain operator insertions in a conformal theory \cite{[Sa],[S35]}.
The parameters $n_j$ will then
provide a necessary freedom allowing us to
relate the \V\ \cs\ \req{Lontau}
to {\it\de s\/} corresponding to null states in
the conformal theory.
The relation with the decoupling equations
is a crucial step \cite{[S35]} that will allow us
to go beyond the standard `fermionic'/Grassmannian
construction \cite{[DDKM]} for the tau function\footnote{
See also ref. \cite{[KSch]} in which solutions to
string equations were constructed in the Grassmannian
language.}.
The \de s are a primary tool in the
analysis of $d\!<\!1$ models, and the study of the corresponding
null vectors has a history of its own
\cite{[FF]}--\cite{[FGPP]}; the
relevant structures are known to possess in certain
cases relations to other problems in physics and mathematics
\cite{[GoSi],[GoSe],[P]}.
It will be amusing to find that
the null vectors incorporate
the set of $p\!\geq\!-1$ \V\ \cs\ (obviously, the \V\ generators
$\sL_p$ are quite distinct from the \V\ generators
represented on a CFT).

More specifically, the \K-\M\ transform relates the \V\ \cs\ to
minimal models that
turn out to be {\it dressed\/} with an extra scalar
that plays a similar r\^ole to that
of the Liouville field in the formalism of \cite{[Da],[DK]}. The
Miwa parameters $n_j$ then acquire the meaning of
the corresponding $U(1)$ `Liouville' charges.
The correspondence is achieved via
an ansatz \cite{[S35]} presenting the
(Virasoro-constrained!) tau function as a correlator of a
product of certain dressed primary field operators in a
minimal model: For any value of the index $i$ chosen,
the constraint

\BE\sump\!z_i^{-p-2}\sL_p\tau=0\label{zconstraints}\EE

\noi
determines
the dependence of $\tau(t)\equiv\tau\{z\}$ on $z_i$ via

\BE\tau=\Bigl<\ldots\Psi(z_i)\ldots\Bigr>\,,\label{firstansatz}\EE

\noi
where $\Psi$ is a dressed $(l,1)$ or $(1,l)$
primary state (thus possessing a null descendant
at level $l$), and the dots denote insertions at the other
points $z_{j\neq i}$ of any of the $(l',1)$ or  $(1,l')$ dressed
primary fields. Let us concentrate on just the one
at $z_i$.
The  respective Miwa parameter $n_i$,
which becomes the `Liouville' $U(1)$ charge of $\Psi$,
must be related to
the `spin' $\sJ$ parametrizing
the \cs\ via

\BE2\sJ-1={l-1\over n_i}-{2n_i\over l-1}~.\label{Jnl}\EE

\noi
This illustrates the importance of introducing parameters into
both the generators
\req{Lontau} and the ansatz \req{Miwatransform} for the KP times.
Moreover,
we find that {\it all\/} the \de s corresponding to
each of the other primary fields in the correlator
\req{firstansatz}
are satisfied by virtue
of {\sl the same\/} set of \V\ \cs, although the corresponding
values of $l'$ and $n_{i'}$ would in general differ from those
on the RHS of \req{Jnl}.

Reversing the argument, we can say
that the correlators \req{firstansatz} {\it solve\/} the \V\ \cs.
This result can be considered as a variation of
a general conclusion, drawn from Matrix Models, that
\V-constrained integrable \hs\ {\it are\/} related to minimal
matter interacting with topological 2$d$ gravity\footnote{
There is, in fact, a certain gap in the argument, as
we do not actually prove
the (Miwa-transformed) Hirota bilinear equations for the RHS of
\req{firstansatz}. What we can nevertheless show is that the
\de s satisfied by the correlator in \req{firstansatz}
are consistent with the KP
evolutions, which is a strong
supporting argument in favour of the ansatz \req{firstansatz}
for the tau function.}.
An attempt towards understanding of the ansatz \req{firstansatz} in
more general terms is given in the next subsection.

\subsection{Solving Virasoro constraints
by a `target-space' theory}\lvm
The approach pursued in this paper is in a certain sense `dual' to the
usual way the time variables $t_r$ enter the theory. Recall
\cite{[K],[W2]} that these are used as `sources' or, `external'
parameters, to build up a generating function for the `string'
correlation functions in a topological gravity\,+\,matter
theory:

\BE\log\tau(t)\sim\sum_h\!\sum_N\!\sum_{\{p_a\}}\L\langle
\phi_{p_1}\ldots\phi_{p_N}\R\rangle_h t_{p_1}\ldots t_{p_N}
\label{W}
\EE

\noi
(a formula of this type is actually valid for a KdV tau function,
but we ignore such differences in this, very qualitative, discussion).
On the other hand, the formulation which we develop below
expresses the tau function as a {\it single\/} correlation function
in what thus becomes an `effective' theory in the sense that its
genus-zero correlators provide an exact solution to the \V\ \cs\ and thus
account for the sum
over all the genera of the `string' correlation functions.
The theory on the RHS of \req{firstansatz}
might be thought of as living
``in the $t$-space", or, in other words, the appropriate
two-dimensional
space is constructed out of the times/coupling
constants/deformation parameters\footnote{
Note also the recent observations \cite{[CV],[Du]}
that the dependence of the structure functions
on the deformation parameters in topological theories is
governed by the same equations as those
describing a space-time dependence of certain non-linear
soluble systems.
}.

Although the worldsheet correlators involved in eq.~\req{W}
are those of a topological theory, we may try to understand
the ansatz \req{firstansatz} qualitatively
by considering for instance a more familiar
object, (the generating function for) the tachyon
scattering amplitudes summed over all genera \cite{[FT]}:

\BE\new\BA{rcl}
\tau[T]&=&\sum_{h\geq0}g^{2(h-1)}\L\langle
e^{\int\!d^2\zeta\,T(X(\zeta))}\R\rangle_h\\
{}&=&
\sum_{h\geq0}g^{2(h-1)}\sum_N{1\over N!}
\int\!d^Dp_1\ldots\int\!d^Dp_N
\L\langle\prod_{a=1}^N\int\!d^2\zeta_a
e^{ip_aX(\zeta_a)}\R\rangle_h\tT(p_1)\ldots\tT(p_N)\,,\EA
\label{tau(T)}\EE

\noi
where the brackets refer to a path integral in the theory with
the free action
$$ S=\int\!d^2\zeta\,\d X\bar{\d}X\,,$$
with the target-space indices suppressed (so, $pX=p_\mu X^\mu$,
$\mu=1,\ldots,D$, etc.). The tachyon field $T$ thus plays the r\^ole
of a set of `external parameters'.
Comparing with the ansatz \req{firstansatz},
we see that for the (hypothetical)
`topological analogue' of \req{tau(T)} for $D\leq2$
it would be possible to perform the sum over all the genera, which
would suggest the `effective' identifications
$$ t_r\sim\tT(p)\,.$$

Moreover, the tree {\sl effective action\/} corresponding to the theory
\req{tau(T)}, which is given as an integral over the $D$-dimensional
target space, can be considered for $D\!=\!2$ as defining another
{\it conformal invariant\/} theory. For $D\!=\!2$ (the $d\!=\!1$ string
theory) there is just one propagating mode, the massless tachyon,
and it is thus plausible that the quantum string theory
loop expansion can be effectively represented by a quantum
{\it field\/} theory,
which (since $D\!=\!2$ and the metric is also quantized)
can in fact be a conformal theory.
The $\Psi_j$ from
\req{firstansatz} are then to be thought of as operators
of this theory, which must in principle be determined by
the effective action\footnote{
We are grateful to R.~Dijkgraaf and A.~Tseytlin
for important discussions of this point.}.
Then, the ansatz \req{firstansatz} suggests
that an appropriate `topological version'
of the above argument is apparently true for $d\!<\!1$
as well, with the
{\it effective\/} theory being just the dressed minimal
model (in fact, as we discuss later, essentially an $N\!=\!2$
minimal model!).

Also for $d=1$, the way the space-time dependence (i.e.\ that on the
$z_j$) emerges can probably be considered along the lines of ref.
\cite{[DMP]}, where {\it two\/} sets of times appeared,
$t_r$ and ${\bar{t}}_{r}$, which leaves a possibility to express
one of them via the $z_j$, as in \req{Miwatransform}, while
viewing the other set as `external' parameters.
As a related problem, let us note that of the correspondence
between the ansatz \req{firstansatz} and the standard
`fermionic' representation for the tau function:
while
the fermionic/Grassmannian construction of ref.\ \cite{[DDKM]}
applies to a general tau function, eq.\ \req{firstansatz}
necessarily gives a solution to the Virasoro \cs\
(and, at the same time, refers to a particular
central charge-$d$ model).
It may be expected that, for $d=1$ at least,
bosonizing the matter (which would leave us with
just two, the matter and the `Liouville', scalars)
would allow a `decoupling' of
one of the two scalars and refermionizing of
the remaining one into the
`standard' fermions.

\subsection{Decoupling equations as BRST invariance}\lvm
Given a minimal model and wishing to relate it to the \V\ \cs,
we would have to dress
it with an extra scalar in a particular way, called the
`\K-\M' dressing prescription,
which would then allow us to interpret the $U(1)$ charges as
the Miwa parameters and eventually recover the \V\ \cs\ from
the \de s. An important point is that
these latter must also be chosen in a special way
from a family parametrized by the $U(1)$ charge of the
primary field\footnote{This choice does not mean any
loss of generality; it specifies a particular `lifting'
of an arbitrary purely-matter null vector into
a null vector in the tensor product theory.}.
These particular recipes, which make
the \de s amenable to the \K-\M\ transform,
are in fact
inherited from BRST invariance in a topological
theory.
Namely, the `\K-\M' dressing prescription
follows from the conditions defining chiral primary
states \cite{[LVW]} of the twisted
$N\!=\!2$ (`{\sl topological\/}') algebra \cite{[W-top],[EY]}.
In addition, the particular \de\ to be chosen among
a family of \de s parametrized by the `Liouville' charge, is
determined just by the BRST-invariance condition
in the topological algebra.
Thus `{\it the\/}' \de s that are related to Virasoro \cs\
are essentially those of the $N\!=\!2$ model.
We employ here the fact that
any $d\!\leq\!1\bigcup d\!\geq\!25$
matter can be embedded into a topological (twisted $N\!=\!2$)
theory, according to a construction
\cite{[GS2]} of the topological
algebra
generators in terms of matter~+~`Liouville'~+~$bc$~system fields.
Unlike the case with the usual reparametrization
ghosts, the $b$ field has conformal dimension 1~\footnote{
So that, {\it if\/} one chooses to
bosonize the matter, one would have
two scalars with the opposite signatures and spin-$(1,0)$ ghosts,
which constitute a topological sigma model in a flat
$2D$ target space.}.
We call this the `mirror' BRST construction,
in contrast with
another one involving $c\!=\!-26$ reparametrization
ghosts (see below).
Then, the condition that the BRST-exact states be factored out
from the representation of the topological algebra, boils down,
in the matter\,$+$\,`Liouville' sector, to the `\K-\M'-dressed
\de s (\ie\ precisely those \de s that allow
the \K-\M\ transform and thereby lead to the \V\ \cs).

On the other hand, exploiting the other
realization \cite{[GS2]}
of the topological algebra, which
involves {\it spin-2\/}
$c\!=\!-26$ ghosts, allows us
to recover the basic features of the DDK formalism.
This can be summarized in a diagram,

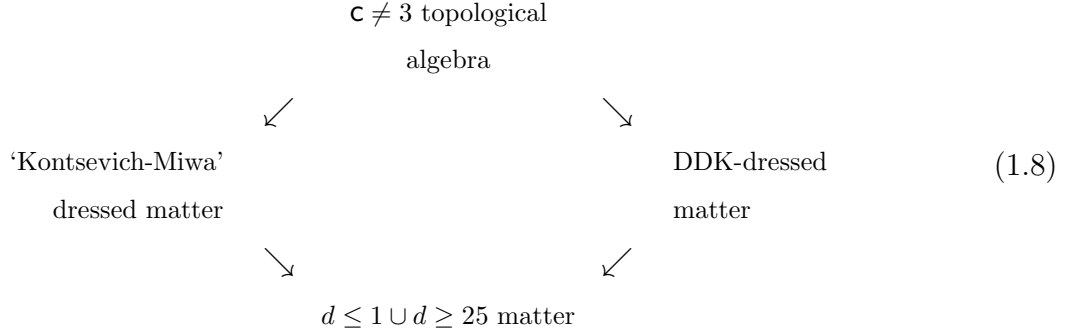
\BE\new\BA{rlcrl}{}&{}&\mbox{{\footnotesize $\BA{c}\ctop\neq3\ {\mathrm
{topological}}\\
{\mathrm{algebra}}\EA $}}&{}&{}\\ {}&\swarrow&{}&\searrow&{}\\
\mbox{{\footnotesize $\BA{r}{\mathrm
{`\K\mbox{-}\M\mbox{'}}}\\ {\mathrm{dressed\ matter}}\EA
$}} &{}&{}&{}&\mbox{{\footnotesize $\BA{l}{\mathrm{DDK\mbox{-}dressed}}\\
{\mathrm{matter}}
\EA $}}\\ {}&\searrow&{}&\swarrow&{}\\ {}&{}&\mbox{{\footnotesize
$d\leq1\cup d\geq25$ matter}}&{}&{}\EA\label{diagram}\EE

\noi
which applies to an arbitrary
$d\!\leq\!1\bigcup d\!\geq\!25$
matter theory and thus shows
that any such theory can be embedded into a topological theory
with \tcc\ (or, the `anomaly')
$\ctop\!\neq\!3$~\footnote{
The issue of interpreting matter theories together with the corresponding
ghosts as topological theories was addressed probably for the first time
in \cite{[DVV-pr]}.
The relation \cite{[GS2]} between
the twisted $N\!=\!2$ theories and the ordinary matter, as given by
the two right arrows in \req{diagram}, has recently
received in \cite{[BLNW]} a powerful
generalization to the case of
W matter. It is a particular case of the
equivalence of categories investigated in \cite{[G]}.}.
The lower arrows tell us that the matter part identified
inside the matter + `Liouville'
theory is of course the same in both cases,
as given by refs.\ \cite{[BPZ],[FQS],[DF]}\footnote{
The null vectors in the {\it matter\/} sector (without
dressing) are of course the standard ones \cite{[FF],[BPZ]}.
Thus the $U(1)$
current might be considered superficial, as soon as it cannot affect the
`dynamical' content of
the story, which can only be based on {\it the\/} \V\
null vectors.
This is, however, precisely what we mean by {\sl dressing\/}:
the r\^ole of the $U(1)$
current is to rearrange the standard \de s so as to
make them amenable to
the \K-\M\ transform; the current does this job
`uniformly' for different-level \de s.}. It appears that only
the `\K-\M'
dressed version is related directly to the integrable formulation,
while the two dressings
should represent `mirror' versions of the same theory;
as we discuss in Sect.~3A, they both result
from the two possible twistings of the proper $N\!=\!2$ algebra.

The matter central charge $d$ is given by

\BE d={(\ctop+1)(\ctop+6)\over\ctop-3}\label{d(c)}\EE

\noi
in terms of the \tcc\ $\ctop$, and therefore the restrictions
$d\!\leq\!1$ or $d\!\geq\!25$
might be viewed
as a result of the `breakdown' of the
twisted $N\!=\!2$ symmetry.
The standard formula for the dimension
of the $(l,1)$ highest-weight states
then follows from the $N\!=\!2$ machinery as
\BE\delta^{(l)}
=-{(l^2-1)\over4}\L({\ctop-3\over6}\R)^{\!\pm1}+{1-l\over2}
={13-d\pm\sqrt{(1-d)(25-d)}\over48}(l^2-1)+{1-l\over2}\,.
\label{deltagen}\end{equation}
Note also that
as $\ctop$ grows from $-\infty$ to 3 and from 3 to
$+\infty$,
each of the allowed values of $d$ is taken twice, except for the
extrema of the function \req{d(c)},
$(\ctop\!=\!-3,d\!=\!1)$ and $(\ctop\!=\!9,d\!=\!25)$. Thus the \tcc\
provides a `two-sheeted covering'
of the allowed region of the matter \cc.

By \tcc\
we mean here the parameter
(the true \cc\ of the untwisted $N\!=\!2$ algebra
\cite{[W-top],[EY]})
appearing in the topological algebra

\BE\new\BA{lclclcl}
\L[\cL_m,\cL_n\R]&=&(m-n)\cL_{m+n}\,,&\qquad&[\cH_m,\cH_n]&=
&{\ctop\over3}m\Kr{m+n}\,,\\
\L[\cL_m,\cG_n\R]&=&(m-n)\cG_{m+n}\,,&\qquad&[\cH_m,\cG_n]&=&\cG_{m+n}\,,
\\
\L[\cL_m,\cQ_n\R]&=&-n\cQ_{m+n}\,,&\qquad&[\cH_m,\cQ_n]&=&-\cQ_{m+n}\,,\\
\L[\cL_m,\cH_n\R]&=&\multicolumn{5}{l}{-n\cH_{m+n}+{\ctop\over6}(m^2+m)
\Kr{m+n}\,,}\\
\L\{\cG_m,\cQ_n\R\}&=&\multicolumn{5}{l}{2\cL_{m+n}-2n\cH_{m+n}+
{\ctop\over3}(m^2+m)\Kr{m+n}\,,}\EA\qquad m,~n\in\oZ\,.\label{topalgebra}
\EE

We thus establish a path leading
from \req{topalgebra} to the \V\ \cs\ \req{Lontau}.
The relation between, say, $\cL_n$ and $\sL_p$ appears not quite
obvious. In particular, we will see that
(for $\ctop<3$, for instance)
the respective parameters are related via
$\sJ=\half+\half\sqrt{{3-\ctop\over3}}-\sqrt{{3\over3-\ctop}}$\,.
We will proceed in two steps, by first reducing from the
BRST-exact topological states
to the `\K-\M' dressed \de s, and then extracting
from the latter the \V\ \cs.

The fact that it is essentially the $N\!=\!2$ symmetry that leads to both
the DDK formalism and the \V\ \cs,
suggests regarding it as a
certain unifying notion. In a worldsheet phase the
$N\!=\!2$ generators split into {\it reparametrization\/} ghosts
plus matter (including the
Liouville), while a different splitting involving $c=-2$ ghosts
(cf.\ \cite{[Di]})
is related to the `integrable'
formulation. The two pictures differ by choosing one out of the two
possible twistings of the proper $N\!=\!2$ algebra \cite{[Dprc]}.

It was first observed in
\cite{[S35]} that the level-2 \de s essentially coincide
with the \V\ \cs\ \req{zconstraints} in which the
\bc\ $2\sJ-1$ is given by eq.\ \req{Jnl} for $l\!=\!2$.
The level-$l$ generalization, as given by \req{Jnl},
was further suggested in \cite{[GS]}. However, the interpretation
given there of the \K-\M\ transformed level-3 \de\
requires being corrected. The actual mechanism
underlying the correspondence between \V\ \cs\ and the \de s
has proved to consist in the
{\it factorization\/} of the \de s through the \V\ generators
\req{Lontau} with the appropriate \bc\ \req{Jnl}.

\bigskip

This paper is organized as follows. In Sect.~2, we review the issue of
the \V-constrained KP
\h\ and give a preliminary analysis, to be considered as a
motivation
(or, as an `elementary explanation') for the subsequent
use of the \K-\M\ transform, which
relates the
`integrable' formalism to the conformal field-theoretic data. As
a `toy' case, we subject the tau function to only the $\sL_{\pm1}$ and
$\sL_0$ \cs, and show that these are precisely the projective Ward
identities of the corresponding conformal theory,
provided the parameters involved satisfy certain relations.
Starting with Sect.~3, we adopt as a `first principle' the notion of
twisted $N\!=\!2$
symmetry and show how the associated BRST-invariance condition
gives rise to exactly the \de s
that turn into the \V\ \cs\ via the \K-\M\ transform.
To `legitimate' taking the topological algebra as a starting point,
we show in the Appendix to Sect.~3 that the other
reduction of the topological algebra gives
the standard DDK formalism.
The dressed \de s that follow from
the BRST-invariance condition
for ghost-independent correlators
can also be constructed directly, by imposing
certain restrictions on the general tensor product null vectors.
This is considered in Sect.~4.
Further, in
Sect.~5 we show that the \K-\M\ transform leads to a reformulation
of the \de s in terms
of \V\ \cs. In Sect.~6, we suggest a generalization of
the lowest-$l$ cases
to arbitrary $l$.
Sect.~7 contains several
concluding remarks and an outlook.

\section{Constrained KP \h\ and \hfill\break
the \K-\M\ transform}
\subsection{KP hierarchy}\lvm
Let us start with
the KP \h\ \cite{[DDKM]}. It can be described either in the evolutionary
form, as an infinite set of equations on (coefficients of) a
pseudodifferential operator, or as (Hirota) bilinear relations on the tau
function. The tau-functional description may be considered
as having the more
direct relevance to `physics', being related to the partition
function, while
the
evolutionary form has all the usual advantages due to the introduction of
a spectral parameter and the associated wave function. The wave function
depends on the spectral parameter $z$ via

\BE\new\BA{c}\psi(t,z)\equiv
e^{\xi(t,z)}w(t,z)=e^{\xi(t,z)}~{\tau(t-[z^{-1}])\over\tau(t)}~,\\
\xi(t,z)=xz+\sumr t_rz^r\EA\label{w}\EE

\noi
where

\BE t\pm [z^{-1}]=\L(
t_1\pm z^{-1},\,t_2\pm\half z^{-2},\,t_3\pm {1\over 3}
z^{-3},\,\ldots\R)\,.\label{timesshift}\EE

\noi
(the plus signs are encountered in the adjoint wave function). Here
$t\!=\!(t_1,t_2,t_3,\ldots)$
are the time parameters of the \h. Knowing the
wave function $w(t,z)$ allows
us to construct the dressing operator $K$ as

\BE w(t,z)=1+\sum_{n\geq1}w_n(t)z^{-n}\quad\Longrightarrow\quad
K=1+\sum_{n\geq1}w_n(t)D^{-n},\qquad D\equiv\dd{x}~.\EE

\noi
Now the evolution equations on $K$ read

\BE\dd{t_r}K=-(KD^rK^{-1})_-K,\qquad r\geq1\label{flows}\EE

The above
form of $z$ dependence, eq.\ \req{timesshift}, is `simulated' by
the Kon\-tse\-vich-Mi\-wa pa\-ra\-me\-tri\-zation \req{Miwatransform}
for the times $t_r$; its heuristic
similarity to
\req{timesshift} may be considered as an `explanation' of its viability.
Below, the \M\ parameters $n_j$
will acquire the r\^ole of `Liouville' $U(1)$ charges
of field operator insertions sitting at the points $z_j$.

\subsection{Virasoro constraints}\lvm
As was noted
in the Introduction, we will consider, on the KP \h,
more general \cs\ than those
that have actually been derived from specific {\it
matrix\/} models, by allowing the \cs\ to depend on a parameter $\sJ$.
That is, we introduce the \cs\

\BE\sL_p\tau=0,\quad p\geq-1\,,\label{Lntau0}\EE

\noi
with the \V\ generators $\sL_p$ as
given in eq.\ \req{Lontau}. The parameter $\sJ$, which
would have
parametrized the central charge as $-2(6\sJ^2-6\sJ+1)$, had
the
$\sL_{\leq-2}$ generators been involved as well, can be thought of as the
`spin' (dimension)
of an abstract $bc$ system underlying the \V\ generators,
$\sum_{n\in{\open Z}}{\Sc L}_nz^{-n-2}\sim (1-\sJ)\d bc-\sJ b\d c$.

That the action on
the KP \h\ via the generators \req{Lontau} is compatible
with the KP flows
has been proved in \cite{[S10]} (see also \cite{[GO]} and
references therein). Indeed, when the infinitesimal action
$\tau\mapsto\sL_n\tau$
is translated into an action on dressing operators, it
becomes

\BE\new\BA{c}K\mapsto{{\fr L}}_pK\equiv
\L(K\bigl(\sJ(p+1)D^p+PD^{p+1}\bigr)K^{-1}\R)_-K,\\ P\equiv x+\sum_{r\geq
1}rt_r D^{r-1}=x+t_1+2t_2D+\ldots\EA\label{frLn}\EE

\noi
It is now completely straightforward to check \cite{[S10]}
the compatibility of this
action
with the flows \req{flows},
which holds irrespectively of the value of $\sJ$.
Hence follows, in particular, the
consistency of the \h\ flows
with the \V\ {\it\cs\/}.

Note that eq. \req{Miwatransform} allows us to rewrite the \V\ generators
${\fr L}_p$ as \BE{{\fr
L}}_p=\biggl(K\Bigl(\sJ(p+1)D^p+xD^{p+1}+\sum_j{n_j\over
z_j-D}D^{p+1}\Bigr)K^{-1}\biggr)_-\label{Lpdiff}\EE

\noi
and therefore the constraint

\BE\sump{{\fr L}}_pz^{-p-2}=0\EE

\noi
takes a rather suggestive form\footnote{
{}From the results of \cite{[DVV]}
we know that the quasiclassical (`dispersionless' \cite{[Kr],[Du1]})
limit of
\req{Lpdiff} or \req{LpdiffMiwa} would give a reformulation
of the Virasoro \cs\ in the Landau-Ginzburg theory. Then, by the
construction of the present paper, which relates the \V\ \cs\ to
the (essentially $N\!=\!2$)
\de s, the resulting `quasiclassical' equation should be
a Landau-Ginzburg reformulation of
the \de s of a (twisted) $N\!=\!2$ model.}

\BE\biggl(K\Bigl(-\sJ\dd{z}+x+\sum_j{n_j\over z_j-D}\Bigr){1\over
z-D}K^{-1}\biggr)_-=0\,.\label{LpdiffMiwa}\EE

\subsection{Why the \K-\M\ transform?}\lvm
It will be shown in the subsequent sections
that the \V\ \cs\ \req{Lntau0} can be solved
by substituting for the tau function, considered
as a function of the $z_j$, the ansatz

\BE\tau\{z_j\}=\lim_{N\rightarrow\infty}\langle\Psi_1
(z_1)\ldots\Psi_N(z_N)\rangle\label{ansatz}\EE

\noi
with $\langle~~\rangle$ and the
$\Psi_j$ being, respectively, the chiral
correlation function and primary field operators in a conformal field
theory on the $z$ plane.
These CFT ingredients will be constructed systematically in the subsequent
sections,
while now we would like to consider, as a motivation, a `toy' case
involving only three \cs, $\sL_{-1}$, $\sL_0$ and $\sL_1$. This
would already
allow us to see how the standard CFT
notions can enter the game.

The $\Psi_j$ operators will be identified
as those of a minimal conformal model
tensored
with an extra $U(1)$ current $I$. The r\^ole and the origin of the
current will be discussed below in some detail, but all we need to know at
the moment
is that the \V\ generators can be chosen in such a way that
there is a zero \bc\ in the $U(1)$ sector, and therefore
the correlators of $\exp\b_j\phi(z_j)$, $\phi(z)\sim\int^z\!I$, can be
evaluated easily.
Moreover, it turns out that the exponents $\b_j$ required
by the dressing prescription must be equal to the $n_j$. Thus,

\BE\biggl\langle\prod_j\Psi_j(z_j)\biggr\rangle=
\prod_{k<l}(z_k-z_l)^{-n_kn_l}
\biggl\langle\prod_j\psi_j(z_j)\biggr\rangle\,,\EE

\noi
where the $\psi_j$ should pertain to a minimal conformal model. Recall
further the `projective Ward
identities' \cite{[BPZ]}

\BE\new\BA{r}\sum_j\L(z_j^{p+1}\dd{z_j}+(p+1)z_j^p\delta_j\R)
\biggl\langle\prod_j\psi_j(z_j)\biggr\rangle=0\,,\\
p=-1,0,1\,,\EA\label{projective}\EE

\noi
where $\delta_j$ is the conformal dimension of $\psi_j$.
This gives \BE\new\BA{r}\sum_j\L(z_j^{p+1}\dd{z_j}+ \half\sum_{k\neq
j}n_jn_k{z_j^{p+1}-z_k^{p+1}\over z_j-z_k}+
(p+1)z_j^p\delta_j\R)\biggl\langle\prod_j\Psi_j(z_j)\biggr\rangle=0\,,\\
p=-1,0,1\,.\EA\label{projectivePsi}\EE

It turns out that these equations
are nothing but the three \V\ \cs\ if we use the
\K-\M\ parametrization
\req{Miwatransform}~\footnote{The fact that the
operators on the LHS of eqs.\
\req{projectivePsi} form {\it an\/} $sl_2$ algebra is of course quite
trivial;
less obvious, however, is that transforming the `projective'
$sl(2)$ generators into the time variables allows us to anticipate the
relations, such as eq.\ \req{preliminary},
between the different parameters, to be
derived systematically below from the
analysis of the \de s.}:
For $p=-1$, quite simply, the operator in \req{projectivePsi}
is just

\BE\sum_j\dd{z_j}=-\sum_jn_j\sumr
z_j^{-r-1}\dd{t_r}=-\sumr(r+1)t_{r+1}\dd{t_r}\equiv-\sL_{-1}\,.\EE

\noi
Similarly, for $p=0$ we get

\BE\new\BA{l}-\sum_jz_jn_j\sumr z_j^{-r-1}\dd{t_r}+\half\sum_j\!
\sum_{k\neq
j}n_jn_k+\sum_j\delta_j\\ =-\sumr
rt_r\dd{t_r}+\half\biggl(\sum_jn_j\biggr)^2+\sum_j\biggl(\delta_j-
{n_j^2\over2}\biggr)\\ =-\sumr rt_r\dd{t_r}\equiv-\sL_0\,.\EA\EE

\noi
The `unwanted' terms here have been cancelled by imposing the relation

\BE\delta_j=\half n_j^2-\half Qn_j\label{preliminary}\EE

\noi
with $Q=\sum_jn_j$ (we will derive these relations
in a more systematic way in
\req{nequation};
for a given $\delta_j$, eq.~\req{preliminary}
allows us to determine the
corresponding $n_j$).
And finally, for $p=1$ we obtain the operator

\BE\new\BA{l}
-\sum_jn_j\sum_{r\geq0}z_j^{-r}\dd{t_{r+1}}+\half\sum_j\!\sum_{k\neq
j}n_jn_k(z_j+z_k)+2\sum_jz_j\delta_j\\ =-\sum_jn_j\dd{t_1}-\sumr
rt_r\dd{t_{r+1}}+\sum_jn_jz_j(Q-n_j)+ 2\sum_jz_j\delta_j\\ =-\sumr
rt_r\dd{t_{r+1}}-Q\dd{t_1}\equiv-\sL_1\,,\EA\EE

\noi
from which we see that $Q$ takes the r\^ole of the \bc\ underlying the \V\
generators \req{Lontau},

\BE Q=\sQ\equiv2\sJ-1~.\label{QsQ}\EE

The above
should be viewed merely as a motivation for using the \K-\M\
transform and the ansatz \req{ansatz} for the tau function.
It is significant, however, that
already such a generic property of conformal models as the projective
invariance,
is captured by the standard $\sL_{\pm1}$, $\sL_0$ \cs\ in terms
of the times $t_r$, {\it with the parameters involved
being related as in eqs.\ \req{preliminary},\,\req{QsQ}}.

The rest
of the Virasoro constraints $\sL_{\geq2}$ no longer follow from the
projective invariance of CFT, but rather from the \de s, which contain the
dynamical information about a particular minimal model.
The actual derivation of the Virasoro constraints from the \de s
will be considered in Sect.~5.

\bigskip

Searching for a
most systematic approach, we will switch in the next section
to the topological algebra.
It is from the BRST invariance
in its highest-weight representations
that we will be able to recover at the end the (specially dressed)
\de s, and then the \V\ \cs\ \req{Lntau0}.
The above relations \req{preliminary}
and \req{QsQ} will reappear in the course of the
derivation.

\section{From the twisted $N=2$ and
BRST invariance\hfill\break
to dressed null states}\label{spin1section}
\lvm We start in this section with a construction of the topological
algebra
\req{topalgebra}, and then use this construction to `reduce' the
requirement of BRST invariance, taken
in the form of factoring out BRST-trivial states,
to the dressed \de s.
The construction of the algebra itself is level-independent,
while the reduction has to be considered level by level.
To legitimate taking the topological
algebra as our starting point, we show in the Appendix to this section
that the $N\!=\!2$ symmetry may
claim the rights of the DDK formulation: in fact,
the topological algebra
admits a reduction to the DDK formalism
and thus can be viewed as a generalization of it.
We also comment in the Appendix
on the relation between the two constructions
of the topological algebra. Only one of these
will be used in the rest of the paper, and it
will eventually lead us, via the \K-\M\ transform, to the
\V\ \cs.

\subsection{A `mirror' BRST
construction for the topological algebra}\lvm
The fields used
to construct our specific realization of
the topological algebra include
matter,
an additional $U(1)$ current $I$, which will be referred to as the
`Liouville' current, and a $c=-2$
$bc$ system. This spin-1
(anticommuting) $bc$ system (the $\eta\xi$ system, \cite{[FMS]})
is defined by

\BE\new\BA{l}b(z)=\sum_{n\in{\open Z}}b_nz^{-n-1}~,\qquad
c(z)=\sum_{n\in{\open Z}}c_nz^{-n}~,\\ \{b_n,c_m\}=\Kr{m+n}~,\qquad
b_{\geq0}\ket0_{{\mathrm{gh}}}=c_{>0}\ket0_{{\mathrm{gh}}}
=0~.\EA\label{spin1}\EE

The matter \V\ generators
combine with the Sugawara `Liouville' contribution \hfill\break
${}-\half\sum_{n\in\oZ}\!:\!I_{m-n}I_n\!:$
into the generators $L_m$ which satisfy

\BE\new\BA{rcl}
\L[L_m,L_n\R]&=&(m-n)L_{m+n}+{d+1\over 12}(m^3-m)\delta_{m+n,0}\,,\\
\L[L_m,I_n\R]&=&-nI_{m+n}\,,\\
\L[I_m,I_n\R]&=&-m\Kr{m+n}
\EA\label{thetheory}\EE

\noi
where the 1 in the \cc\ $d+1$ is the $U(1)$ contribution.
We find it
convenient to `twist' the \V\ generators as

\BE\hL_m=L_m+\half Q(m+1)I_m\,,\label{hatshift}\EE

\noi
where $Q$ is the {\it matter\/} \bc\

\BE Q=\sqrt{{1-d\over3}}\,,\label{Q(d)}\EE

\noi
which therefore becomes also that of the `Liouville' scalar.
Then, the \cc\ of the new \V\ generators is just
$1-3Q_{{\mathrm{m}}}^2+1+3Q_{{\mathrm{L}}}^2=2$, so that the
corresponding formulae in \req{thetheory} get replaced by \BE\new\BA{rcl}
\L[\R.\!\hL_m,\hL_n]&=&(m-n)\hL_{m+n}+{2\over 12}(m^3 -m)\Kr{m+n}~,\\
\L[\R.\!\hL_m,I_n]&=&-nI_{m+n}-\half Q(m^2+m)\Kr{m+n}~.\EA\label{hat}\EE

Now, centreless \V\
generators as those in \req{topalgebra} can be constructed by adding the
$c=-2$ ghost contribution:

\BE\cL_m=\hL_m+l_m,\quad
l_m=\sum_{n\in{\oZ}}n\!:\!b_{m-n}c_n\!: \label{L}\EE

\noi
[the ghost \emt\ being $T_{{\mathrm{
gh}}}(z)=-\!:\!b\d c\!:\!(z)$]. Further,
introducing the ghost current $i=-:\!bc\!:$\,, we define the topological
$U(1)$ current as

\BE\cH_m=i_m+\sqrt{{3-\ctop\over3}}I_m\,,\label{H}\EE

\noi
so that

\BE[\cH_m,\cH_n]={\ctop\over3}m\Kr{m+n}\,.\EE

\noi
It follows that the topological algebra commutator

\BE[\cL_m,\cH_n]=-n\cH_{m+n}+{\ctop\over6}(m^2+m)\Kr{m+n}\label{LH}\EE

\noi
can now be established {\it provided
the matter \cc\ $d$ is related to the \tcc\ $\ctop$ by eq.~\req{d(c)}.}

Next, we need to construct the remaining generators
of the topological algebra
\req{topalgebra}.
As the ghost field $b$ is of dimension 1, the
ansatz for
the BRST current $\cQ\sim cT$, which works for $c$ being of
dimension $-1$ (see the Appendix to this section)
does not apply here. Instead, we can identify the modes of a
spin-1 odd current $\cQ(z)$ simply as

\BE\cQ_m=b_m\,.\label{Q}\EE

\noi
On the other hand, it is the spin-2 fermionic field $\cG(z)$ that now
comprises the
`non-trivial' terms, usually characteristic to the BRST generators
when these are built \cite{[FMS]} using a spin-2 $b$ field:

\BE\cG_m=2\sum_{p\in\oZ}c_{m-p}\hL_p+2{}\sqrt{{3-\ctop\over3}}
\sum_{p\in\oZ}(m-p)c_{m-p}I_p
{}+\sum_{p,r\in\oZ}(r-p):\!b_{m-p-r}c_rc_{p}\!:
{}+{}{\ctop\over3}(m^2+m)c_m\,.\label{G}\EE

To avoid misunderstanding, let us stress that
the $\hL_p$ generators here include, besides the
`improvement' term written out explicitly in
\req{hatshift}, also the Sugawara `Liouville'
contribution $-\half I^2$. Thus,

\BE\hL_p=\bar{L}_p-\half\sum_n :\!I_{p-n}I_n\!:{}+{}\half Q(p+1)I_p
\label{LLbar}\EE

\noi
where the $\bar{L}_p$ are purely-matter Virasoro generators,
with the \cc\ given by \req{d(c)}.

The coefficient in front of
the second term in \req{G} (and the same coefficient in \req{H})
is real for $\ctop<3$;
the regions $\ctop<3$ and $\ctop>3$ are `mirrored'
in that the matter
and the `Liouville' take the place
of each other: all
the coefficients in the above
ansatze for the topological algebra generators
can be kept real when going over from $\ctop<3$ to $\ctop>3$, by reversing
the signature of the fields.

All the commutation relations \req{topalgebra} can now be verified
straightforwardly.
We have thus arrived at a realization of the topological
algebra \req{topalgebra}
and the relation \req{d(c)} between the central charges.
It is valid, irrespective
of whatever $d\!\leq\!1$ or $d\!\geq\!25$
matter theory is taken at the start\footnote{
Note in particular that there is no need to bosonize
the matter; however, when explicitly bosonizing it,
a similar construction (actually, the one discussed in the
Appendix to this section) can also be arrived at via the
hamiltonian reduction \cite{[BLNW]}.}.
Below, we will use this realization in order to translate
the BRST-invariance condition into the matter\,$+$\,`Liouville'
theory.

\subsection{Decoupling equations from BRST invariance}\lvm
Let us
consider a representation of the topological algebra built on a
BRST-invariant chiral primary state $\ket\Phi$ \cite{[LVW]}

\BE\new\BA{rclcrcl} \cQ_0\ket\Phi&=&0,&{}&\cG_0\ket\Phi&=&0,\\
\cL_{\geq0}\ket\Phi&=&\cH_{\geq1}\ket\Phi&=
&\cG_{\geq1}\ket\Phi&=&\cQ_{\geq1}\ket\Phi~=~0,\\
\cH_0\ket\Phi&=&\htop\ket\Phi.&{}&{}&{}&{}\EA\EE

\noi
The topological $U(1)$ charge $\htop$ is thus the only non-zero parameter
that distinguishes between such states.
The dimension zero condition will translate below into the
`\K-\M' dressing condition, while
a `level-by-level' imposition of the
BRST invariance in the above representations
will give rise to the particular
\de s that are amenable to the \K-\M\ transform.
Namely,
provided $\htop$ is related to $\ctop$ by a certain quadratic
equation [see \req{htopctop} for the $(l,1)$ case], there will be
primary states which are BRST-exact and should therefore
be factored out. This is achieved by imposing equations, analogous
to the decoupling ones, which for the ghost-independent
[and hence, in view of \req{Q}, BRST-invariant]
insertions boil down to the `Kontsevich-Miwa' dressed
\de s. These latter
will thus be derived in the remaining part of
this section, while in Sect.~5 their relation
to the \V\ \cs\ will be demonstrated.

\subsubsection{Level-2 reduction and the decoupling equation}\lvm
We start with the simplest, level-2,
case of imposing
the BRST-invariance condition in a representation of the
topological algebra \req{topalgebra},
to show that under the splitting, as in Sect.~3.1, of a
BRST-invariant highest-weight state
into matter\,$\otimes$\,`Liouville'\,$\otimes$\,ghosts,
the matter\,$\otimes$\,`Liouville'
part becomes a null vector dressed in
such a way that it allows a subsequent
application of the \K-\M\ transform.

The BRST-invariant states, $\cQ_0\ket\Xi=0$, at level 2 are of the form

\BE\ket{\Xi}=\L(\a\cL_{-1}^2+\cL_{-2}+\Gamma\cH_{-1}\cL_{-1}+
\half\Gamma\cQ_{-1}\cG_{-1}\R)\ket{\Phi}\,,\label{Xi}\EE

\noi
with $\a$ and $\Gamma$ arbitrary so far. A crucial point
is that the
BRST invariance rules out the $\cH_{-2}$ and $\cH_{-1}^2$ terms.
Later, we will see that
this property, which persists to higher levels,
can be put into the basis of an independent `direct' derivation
of the desired dressed null vectors.

Impose further on $\ket\Xi$ the highest-weight conditions w.r.t. the
topological algebra. From $\cG_1\ket\Xi=0$ we find

\BE2\a+3+\Gamma\L(\htop+1+{\ctop\over3}\R)=0\,.\EE

\noi
Further, the constraint
$\cH_1\ket\Xi=0$ gives two equations

\BE
\L\{\new\BA{l}2\a\htop
+\Gamma\L({\ctop\over3}-1\R)=0\\1+\Gamma\htop=0\EA\R.\EE

All the highest-weight conditions

\BE\cQ_0\ket\Xi=\cQ_{\geq1}\ket\Xi=\cG_{\geq1}\ket\Xi
=\cL_{\geq1}\ket\Xi=\cH_{\geq1}\ket\Xi=0\label{hwconditions}\EE

\noi
are now satisfied. Indeed, the $\cQ_1$-
and $\cG_2$-conditions follow from the established ones
via $\cQ_1=[\cQ_0,\cH_1]$,
$\cG_2=[\cH_1,\cG_1]$.
Further, $\cL_1=\half\{\cG_1,\cQ_0\}$ follows as well,
and therefore, via $[\cL_1,\cG_n]=(1-n)\cG_{n+1}$ and
$[\cL_1,\cQ_n]=-n\cQ_{n+1}$,
so do $\cG_{\geq3}$ and $\cQ_{\geq2}$. Generated
similarly are the $\cH_{\geq2}$ conditions. Finally,
$\{\cG_1,\cQ_1\}=2\cL_2-2\cH_2$,
which implies the $\cL_2$ and hence all the
$\cL_{\geq2}$ conditions.

It follows that

\BE
2\htop^2-\htop\L({\ctop\over3}+1\R)+{\ctop\over3}-1=0\label{hequation}\EE

\noi
{}from which we find two solutions

\BE
\htop=\ccases{\ctop-3\over6}{1}\!,
\qquad\a=\ccases{6\over\ctop-3}{\ctop-3\over6}\!,
\qquad\Gamma=\ccases{6\over3-\ctop}{-1}\!.\label{coefficients}\EE

Now, the state $\ket\Xi$ thus constructed proves to be not only
BRST-invariant, but also BRST-{\it exact\/}:

\BE\ket\Xi=\cQ_0\L(
{\alpha\over2}\cL_{-1}\cG_{-1}+\half\cG_{-2}+\half\Gamma\cH_{-1}\cG_{-1}
\R)\ket\Phi\,.\EE

\noi
Such
states should be factored out, which
can be accomplished by imposing certain equations on the correlation
functions
involving $\Phi$ (in obvious analogy with the use of \de s to
factor away
null vectors from Verma modules \cite{[BPZ],[DF]}).

At this point we introduce
into our scheme the {\sl Miwa parameter\/} $\hn\equiv n_i$
as the `Liouville' charge of $\Phi(z_i)$, according to
\req{H}: we set

\BE\htop=\sqrt{{3-\ctop\over3}}\hn\,.\label{introducen}\EE

\noi
Then from eq.\ \req{hequation}
$\hn$ is given in terms of the \tcc, and hence in terms of the matter
central charge, by

\BE\hn=\ccases{-\half\sqrt{{3-\ctop\over3}}}{\sqrt{{3\over3-\ctop}}},
\quad\Longrightarrow\quad
\hn^2={13-d\pm\sqrt{(1-d)(25-d)}\over24}\,.\label{n2(d)}\EE

Now, to perform the reduction to the matter\,$\otimes$\,`Liouville'
theory, we substitute for the topological
generators the expressions
\req{L}, \req{H}, \req{Q}, and \req{G}, and also take

\BE\ket\Phi=\ket\Psi\otimes\ket0_{{\mathrm{gh}}}\,.\EE

\noi
This form of $\ket\Phi$, with no ghost oscillators involved, means
in particular that the state is BRST invariant, $b_0\ket\Phi=0$.
Then, using \req{spin1} and
\req{coefficients}, we find that [for {\it both\/} the upper and the
lower cases in \req{coefficients}, \req{n2(d)}]:

\BE\ket\Xi=\ket\Upsilon\otimes\ket0_{{\mathrm{gh}}}\,,\EE

\noi
where

\BE\new\BA{rcl}\ket\Upsilon&=&\L(\a\hL_{-1}^2+\hL_{-2}+\g
I_{-1}\hL_{-1}\R)\ket\Psi\\ {}&=&\L(\a L_{-1}^2+L_{-2}+\b I_{-2}+\g
I_{-1}L_{-1}\R)\ket\Psi\\ \a&=&{}-{1\over2\hn^2}\,,\qquad
\b~=~\hn-{1\over2\hn}\,
,\qquad \g~=~{}-{1\over\hn}\,.\EA\label{UUpsilon}\EE

\noi
The two expressions for $\ket\Upsilon$ are related by eq.\
\req{hatshift}, in which $Q=\inv{\hn}-2\hn$, as can be found by
comparing eqs.\ \req{n2(d)} and \req{Q(d)}.

We have thus obtained
a null vector in the matter theory tensored with the `Liouville':
$L_{\geq1}\ket\Upsilon=I_{\geq1}\ket\Upsilon=0$.
To get down to
the bottom of the diagram \req{diagram},
we subtract away from the \V\ generators $L_n$
the Sugawara `Liouville'
contribution,
and in this way recover the usual `minimal' null vector in the
$(d,\delta)$ Verma
module, with the central charge $d$ given by \req{d(c)}.
As for the highest weight $\delta$, it is found to be

\BE\delta=\ccases{-{\ctop\over8}-{1\over8}} {-{\ctop+6\over2(\ctop-3)}}
={5-d\pm\sqrt{(1-d)(25-d)}\over16}\EE

\noi
which is a standard expression.

\bigskip

At the level of
correlation functions, the condition for $\ket\Upsilon$ to
factor out takes the form of a decoupling equation for correlators
of the form

\BE\biggl\langle\Psi(z_i)\prod_{j\neq i}\Psi_j(z_j)
\biggr\rangle\,.\label{corrfunctions}\EE

\noi
Obviously, at least one operator insertion here must be that of $\Psi$,
and thus the corresponding insertion point $z_i$ will from now on be
singled out from the rest of the $z_j$. The other insertions, of
dimensions\footnote{Dimensions refer to those as evaluated in the
matter\,+\,`Liouville'
theory, by operator product expansions with the \emt\ $T(z)=
\sum z^{-n-2}L_n$.}
$\Delta_j$ and $U(1)$ charges $n_j$, may
or may
not coincide with $\Psi$.

In order to derive the \de s from the conditions
$\bigl\langle\Upsilon(z_i)\prod_{j\neq i}\Psi_j(z_j)
\bigr\rangle=0$, we proceed in the standard way \cite{[BPZ]}, but in
addition
restrict
ourselves to a subspace of those operators $\Psi_j$ whose dimensions
and $U(1)$ charges are related by the {\it `\K-\M' dressing
condition\/}:

\BE\Delta_j=-\half Qn_j\stackrel{(l=2)}{=}
(\hn^2-\frac{1}{2}){n_j\over\hn}~.\label{dressing2}\EE

\noi
{}From the point of view of the matter\,+\,`Liouville' theory,
this dressing condition just
says that the
scalar $\phi(z)\sim\int^z\!I$ enters vertex operators with an exponent
determined by the matter part. However, we also know the `invariant'
meaning of \req{dressing2}, which is tantamount to saying that
the $\widehat{\phantom{z}}$-dimension of $\Psi_j$
is zero, and therefore, for ghost-independent insertions,
so is the `topological' $\cL$-dimension; thus the $\Psi_j$
are ghost-independent representatives of chiral primary fields.
In fact, the dressing condition \req{dressing2} is the same as
\req{preliminary}, with $\Delta_j$ the sum of the matter
dimension $\delta_j$ and the Sugawara $U(1)$ contribution $-\half n_j^2$
(and therefore, in view of the general arguments of Sect.~2.3,
the first of the equations in \req{dressing2}
should hold
irrespectively of \de s).

Then, we arrive at a `dressed'
\de, which reads\footnote{This has the form of
a `continuum'
version of the `discrete' {\it master equation\/}
from \cite{[MS]}; see ref.
\cite{[S35]} for a discussion of this point.
The same equation appeared recently
in the solution of the Calogero
model \cite{[BHV]}.}

\BE
\L\{-{1\over 2n_i^2}{\d^2\over\d z_i^2}+{1\over n_i}\sum_{j\neq i}{1\over
z_j-z_i}\L(n_j{\d\over\d z_i}-n_i{\d\over\d z_j}\R)\R\}
\biggl\langle{\Psi}(z_i)\prod_{j\neq i}\Psi_j(z_j)\biggr\rangle=0\,,
\label{mastereq}\EE

\noi
with $\Psi$ being the dressed `21' operator. As we will see
in Sect.~5.1, this equation
is essentially the same as the \V\ \cs\ \req{zconstraints}.

\subsubsection{Level-3 reduction and the decoupling equation}\lvm
The story is much the same for level 3.
The BRST invariance in our realization of the
topological algebra serves as a principle that
fixes the special form of
a null state, which we will recognize below as being
required for the application of the \K-\M\ transform.
That is, for a level-3
topological algebra state

\BE\new\BA{rcl}\ket\Xi&=&(\a \cL_{-1}^3-2\cL_{-2}\cL_{-1}+\b\cH_{-3}+\g
\cH_{-2}\cL_{-1}+\delta\cL_{-3}+\epsilon\cH_{-1}\cL_{-2}\\
{}&{}&{}
+\mu\cH_{-1}^2\cL_{-1}+\nu\cH_{-1}\cL_{-1}^2+\kappa\cH_{-1}\cH_{-2}+\rho
\cH_{-1}^3\\
{}&{}&{}+
a\cQ_{-2}\cG_{-1}+e\cQ_{-1}\cG_{-2}+f\cL_{-1}\cQ_{-1}\cG_{-1}+
g\cH_{-1}\cQ_{-1}\cG_{-1})\ket\Phi\,,\EA\EE

\noi
we first demand that it
be BRST-invariant, $\cQ_0\ket\Xi=0$. It follows that $\rho=0$ and
then $\b=\kappa=0$ (no pure-$\cH$ terms, as was the case for level 2)
along with several other relations. Thus,

\BE\new\BA{rcl}
\ket\Xi&=&(\a\cL_{-1}^3-2\cL_{-2}\cL_{-1}+\g\cH_{-2}\cL_{-1}
+\delta\cL_{-3}+2b\cH_{-1}\cL_{-2}
+g\cH_{-1}^2\cL_{-1}+2f\cH_{-1}\cL_{-1}^2\\
{}&{}&{}+a\cQ_{-2}\cG_{-1}+e\cQ_{-1}\cG_{-2}+f\cL_{-1}\cQ_{-1}\cG_{-1}+
g\cH_{-1}\cQ_{-1}\cG_{-1})\ket\Phi\,,\EA\label{BRSTinv3}\EE

\noi
where $\g=2a+2f-g$.
Further, from $\cH_1\ket\Xi=0$ we find,

\BE\new\BA{rclcrclcrcl}
f&=&{3\over2\htop}~,&{}&\a&=&{3-\ctop\over3\htop^2}~,&{}&a&=&
{\ctop+9-6\htop\over6-2\ctop}~,\\
e&=&{-3\htop\over3-\ctop}~,&{}&g&=&{6\over3-\ctop}~,&{}&\g&=&{\ctop+3-
6\htop\over3-\ctop}+{3\over\htop}~,\\
\delta&=&\multicolumn{9}{l}{-\g\htop~=~{6\htop^2-\htop(\ctop+3)-9
+3\ctop\over3-\ctop}~.}\EA\label{coeff3}\EE

\noi
Then, the condition $\cG_1\ket\Xi=0$ leads to the quadratic equation

\BE\htop^2-{\ctop+3\over3}\htop+{2\over3}(\ctop-3)=0\,,\EE

\noi
whence

\BE\htop=\ccases{{\ctop\over3}-1}{2}.\label{hcases}\EE

Now, we use the same formula \req{introducen} as above to
introduce a Miwa parameter $\hn$ (this formula is independent of the
level and reflects only
the structure \req{H} of the topological current).
Then it follows from \req{coeff3}

\BE\new\BA{rclcrclcrcl}
\a&=&{1\over\hn^2}\,,
&{}&\g&=&\cases{1-{1\over\hn^2}\cr{}-{\hn^2\over2}+\half},
&{}&a&=&\cases{\half+{2\over\hn^2}\cr{}-\half},\\
e&=&\cases{1\cr{}-{\hn^2\over2}},
&{}&g&=&\cases{{2\over\hn^2}\cr{\hn^2\over2}},
&{}&f&=&\cases{{}-{3\over2\hn^2}\cr{3\over4}},\\
\delta&=&\hn^2-1\,.&{}&{}&{}&{}&{}&{}&{}&{}\EA\EE

To see how the matter\,$\otimes\,U(1)$-null vector arises,
consider now that part of (the first seven terms from) the RHS of
\req{BRSTinv3} which contains no ghosts. It reads

\BE\new\BA{rcl}\a\hL_{-1}^3&-&2\hL_{-2}\hL_{-1}+
\g\sqrt{{3-\ctop\over3}}I_{-2}\hL_{-1}+\delta\hL_{-3}\\
{}&+&2e\sqrt{{3-\ctop\over3}}I_{-1}\hL_{-2}
+g{3-\ctop\over3}I_{-1}^2\hL_{-1}
+2f\sqrt{{3-\ctop\over3}}I_{-1}\hL_{-1}^2\label{38}\EA\EE

\noi
We substitute here, as it follows from \req{hcases}
and \req{H}:

\BE\sqrt{{3-\ctop\over3}}=\ccases{-\hn}{{2\over\hn}}\EE

\noi
and act with the operator \req{38} onto the state
$\ket\Psi$, where $\ket\Phi=\ket\Psi\otimes\ket0_{{\mathrm{gh}}}$.
Then, {\it for both the upper and the lower cases\/},
we arrive at the following
{\it null state\/} of the ($L$, $I$) (or
($\hL$, $I$)) algebra:

\BE\new\BA{rcl}\ket\Upsilon&=&\L({1\over
\hn^2}\hL_{-1}^3-2\hL_{-2}\hL_{-1}+{1-\hn^2\over\hn}I_{-2}\hL_{-1}
+(\hn^2-1)\hL_{-3}-2\hn I_{-1}\hL_{-2}\R.\\
{}&{}&\hfill\L.{}+2I_{-1}^2\hL_{-1}
+{3\over\hn}I_{-1}\hL_{-1}^2\R)\ket\Psi\\
{}&=&\L({1\over \hn^2}L_{-1}^3-2L_{-2}L_{-1}+
{3-2\hn^2\over\hn}I_{-2}L_{-1}
+(\hn^2-1)L_{-3}-2\hn I_{-1}L_{-2}\R.\\
{}&{}&\hfill\L.{}+2I_{-1}^2L_{-1}+{3\over\hn}I_{-1}L_{-1}^2+
{(\hn^2-1)(\hn^2-2)\over\hn}I_{-3}+(2-\hn^2)I_{-1}I_{-2}\R)
\ket\Psi\label{UUpsilon3}\EA\EE

\noi
where, as follows from the previous formulae,

\BE\hn^2={13-d\pm\sqrt{(1-d)(25-d)}\over6}\,,\label{nsquared3}\EE

\noi
and

\BE Q={2\over\hn}-\hn\,,\EE

\noi
so that the shift \req{hatshift} takes the form

\BE\hL_m
=L_m+\half(m+1){2-\hn^2\over\hn}I_m\,.
\label{hatshift3}\EE

\noi
The absence of the pure-$I$ terms in the
null state obtained by the action
of $\hL$'s, eq.~\req{UUpsilon3}, is a consequence of the
BRST invariance, which we have seen to suppress the
pure-$\cH$ terms. On the other hand it is just the
condition that the pure-$I$ terms be absent that is
crucial for the application of the \K-\M\ transform to
the corresponding \de. This pattern repeats at higher levels,
as the vanishing of the $I_{-1}^l$ term in the level-$l$
tensor product decoupling equation.

By the above
construction, the `dressed' \de\ corresponding to
the above null vector will be just a particular case of implementing the
BRST-invariance principle ${{\mathrm{Im}}}\cQ\sim0$
for correlators comprising only ghost-free
representatives of chiral primary fields.
The dimension-zero condition, as before, becomes
the `\K-\M' dressing prescription, which now reads

\BE\Delta_j=\half(\hn^2-2){n_j\over\hn}\,.\label{dressing3}\EE

\noi
Then, we arrive at the  \de

\BE\widehat{\cal O}\biggl\langle\Psi(z_i)\prod_{j\neq
i}\Psi_j(z_j)\biggr\rangle=0\label{2}\EE

\noi
where

\BE\new\BA{l}\widehat{\cal O}={1\over \hn^2}{\d^3\over\d z_i^3}
+{}\sumji{1-\hn^2\over (z_j-z_i)^2} \L(\dd{z_j}-{n_j\over
\hn}\dd{z_i}\R)\\{}
+\sumji{1\over z_j-z_i}\L(2 {\d^2\over\d z_j\d z_i}-3{n_j\over
\hn}\ddsc{z_i}\R)-2\sumji\!\sumki{n_k\over(z_j-z_i)(z_k-z_i)}
\L(\hn\dd{z_j}-n_j\dd{z_i}\R)\,.\EA\label{decouplingop3first}\EE

\noi
This decoupling operator is
further studied in Sect.~5, where we show how it can be
related to the \V\ \cs\ in terms of the KP times $t_r$.

\addcontentsline{toc}{subsection}{\protect\numberline{3.A}
{Appendix: DDK formalism from $N\!=\!2$}}

\setcounter{equation}{0}
\def\theequation{\thesection.A\arabic{equation}}

\subsection*{Appendix:
DDK formalism from $N=2$}\lvm
As has been noticed in the Introduction, our starting point,
the
topological algebra, allows us to recover, besides
the `\K-\M'-dressed matter,
also the DDK dressing prescription.
To this end, one should consider, instead of
the ansatz \req{L}, \req{H},
\req{Q}, \req{G}, another reduction of the
topological algebra to matter +
`Liouville' + ghosts \cite{[GS2]}. Namely, let us split away
from the topological generators a spin-2 ghost system. Using the same
notations as in Sect. \ref{spin1section} (although the ghost system is
actually different), we write

\BE \cL_m=\hL_m+l_m,\quad
l_m\equiv\sum_{n\in{\oZ}}(m+n):\!b_{m-n}c_n\!:\label{L26}\EE

\noi
Then the \cc\
read off from the $[\hL_m,\hL_n]$ commutator is 26. Further, the {\it
DDK dressing
prescription\/} \cite{[Da],[DK]}
can be recovered as follows. Recall that for a
spin-$\lambda$ ghost system, the $SL_2$ invariant vacuum state $\ket0$ is
characterized by \cite{[FMS]},

\BE b_{>-\lambda}\ket0=c_{\geq\lambda}\ket0=0\label{cr-an}\EE

\noi
[a particular case of which we have already met in \req{spin1}]. For the
reparametrization ghosts it is thus only the $c_{\geq2}$ out of the $c_n$
modes that annihilate the vacuum, which allows us to split the topological
algebra states as

\BE\ket\Phi=\ket\Psi\otimes c_1\ket0\,.\EE

\noi
In view of \req{L26} and \req{hwconditions}, this implies that the
$\hL$-dimension of
$\Psi$ is $\widehat\Delta=1$. Clearly, the same holds for
all the other chiral primary states $\Phi_j$:
splitting them as

\BE\ket{\Phi_j}=\ket{\Psi_j}\otimes c_1\ket0~,\EE

\noi
we arrive at the conditions

\BE\widehat\Delta_j=1\,,\EE

\noi
which will
fix the DDK dressing prescription as soon as the \bc s are known.
Thus, to complete the derivation, let us first give the expression for the
BRST current modes:

\BE\cQ_m=2\sum_{p\in\oZ}c_{m-p}\hL_p
+\sum_{p,r\in\oZ}(p-r):\!b_{m-p-r}c_pc_r\!:{}+
{}2\sqrt{{3-\ctop\over3}}m
\sum_{p\in\oZ}c_{m-p}I_p+{\ctop\over3}(m^2-m)c_m~.\label{Q26}\EE

\noi
As to the $\hL_p$ generators, note that the comment below eq.~\req{G}
applies in the present case as well.
Now, with the superpartner of the \emt\ being

\BE\cG_m=b_m\,,\label{G26}\EE

\noi
we use the
$\{\cG_m,\cQ_n\}$ commutator to find that the topological current
is given by

\BE\cH_m=\sum_{n\in\oZ}:\!b_{m-n}c_n\!:{}-
{}\sqrt{{3-\ctop\over3}}I_m
\,.\label{H26}\EE

{}From the $[\cL_m,\cH_n]$ commutator we can now derive

\BE [\hL_m,I_n]=-nI_{m+n}-
\sqrt{{3\over3-\ctop}}{\ctop-9\over6}(m^2+m)
\Kr{m+n}\,.\label{hat26}\EE

\noi
The anomaly thus emerging can be expressed as

\BE-
\sqrt{{3\over3-\ctop}}{\ctop-9\over6}=-\half\sqrt{{25-d\over3}}\equiv-
\half Q_{{\mathrm L}}\,,\EE

\noi
where $Q_{{\mathrm L}}$ is the standard \bc\
of the Liouville scalar.
Let us note that the construction \req{L26}, \req{Q26}--\req{H26},
\cite{[GS2]},
for the topological generators was generalized recently to include the
W$_3$ generators \cite{[BLNW]}.

\bigskip

The existence of just two `bosonizations' of the topological
algebra in terms of matter, `Liouville' and ghosts,
follows, according to an observation by Dijkgraaf \cite{[Dprc]},
from the two possible twistings of the proper $N\!=\!2$ algebra.
Both the `spin-1' construction of Sect.~3.1 and the
above `spin-2' version are different twistings of a
`spin-${3\over2}$' construction.
Indeed, consider the $N\!=\!2$ algebra
(we have once again changed the notations, the $\cL_p$ now
being the untwisted generators)

\BE\new\BA{lclclcl}
\L[\cL_m,\cL_n\R]&=&(m-n)\cL_{m+n}+{\ctop\over12}(m^3-m)\Kr{m+n}
\,,&\qquad&[\cH_m,\cH_n]&=
&{\ctop\over3}m\Kr{m+n}\,,\\
\L[\cL_m,\cG_r^\pm
\R]&=&\L({m\over2}-r\R)\cG_{m+r}^\pm
\,,&\qquad&[\cH_m,\cG_r^\pm]&=&\pm\cG_{m+r}^\pm\,,
\\
\L[\cL_m,\cH_n\R]&=&{}-n\cH_{m+n}\\
\L\{\cG_r^-,\cG_s^+\R\}&=&\multicolumn{5}{l}{2\cL_{r+s}-(r-s)\cH_{r+s}+
{\ctop\over3}(r^2-\frac{1}{4})
\Kr{r+s}\,,}\EA\label{N2algebra}
\EE

\noi
where the $\cG^\pm$ have spin $\frac{3}{2}$ and are half-integer
moded. One of the fields $\cG^\pm(z)$ can be identified with the $b$
field of a spin-$\frac{3}{2}$ $bc$ system:

\BE b_r=\cG^-_r\,.\EE

\noi
Let then $c$ be the conjugate, spin-$(-\half)$ field. The $bc$ theory is
defined as before, by $\{b_r,c_s\}\!=\!\Kr{r+s}$
and relations \req{cr-an} for $\lambda\!=\!{3\over2}$; the \emt\
and the ghost current read

\BE l_m=\sum_r\L(r+{m\over2}\R):\!b_{m-r}c_r\!:\,,\qquad
i_m=-\sum_r:\!b_{m-r}c_r\!:\,.\EE

Now, let

\BE Q=-{\ctop+3\over\sqrt{3(3-\ctop)}}\,,\qquad
Q_{{\mathrm{L}}}={\ctop-9\over\sqrt{3(3-\ctop)}}\EE

\noi
(which is real for $\ctop\!<\!3$),
and represent the $U(1)$ current $\cH$ as

\BE\cH_m=i_m-\half(Q_{{\mathrm{L}}}-Q)I_m\,,\EE

\noi
where, as before,
$[I_m,I_n]\!=\!-m\Kr{m+n}$. Similarly, for the Virasoro
generators $L_m$, introduced via

\BE\cL_m=L_m+{1\over4}
(Q_{{\mathrm{L}}}+Q)(m+1)I_m+l_m\,,\EE

\noi
we find the central charge $d+1$ with $d$ given by \req{d(c)},
and therefore, after subtracting the Sugawara
$U(1)$ contribution as in \req{LLbar}, we arrive at
the `matter' Virasoro generators with \cc\ $d$.
We have already seen that,
when reexpressed in terms of $d$, $Q_{{\mathrm{L}}}$
becomes the standard Liouville
\bc\ $\sqrt{{25-d\over3}}$, while $Q$ coincides with the matter \bc\
$\sqrt{{1-d\over3}}$.
Finally, we construct
\BE\cG_r^+=2\sum_{n}c_{r-n}L_n
+\sum_{s,q}(s-q):\!b_{r-s-q}c_sc_q\!:
+\sum_{n}(Q_{{\mathrm{L}}}n+(Q-Q_{{\mathrm{L}}})r
+\frac{1}{2}Q+\frac{1}{2}Q_{{\mathrm{L}}})
c_{r-n}I_n
{}+{}{\ctop\over3}(r^2-\frac{1}{4})c_r\,.\label{G32}\EE

There are, as usual, just {\it two\/} possibilities to twist the algebra
\req{N2algebra}, with either $\cG^+$ or $\cG^-$ acquiring spin 2.
The first twisting is accomplished by setting

\BE\new\BA{rclcrcl}
\cL^{(1)}_m&=&\multicolumn{5}{l}{\cL_m+\half(m+1)\cH_m\,,}\\
c^{(1)}_m&=&c_{m+\half}\,,&\qquad &b^{(1)}_m&=&b_{m-\half}\,,\\
\cH^{(1)}_m&=&\cH_m\,,&{}&{}&{}&{}\\
\cG^{(1)}_m&=&\cG_{m+\half}^+\,,&\qquad &\cQ_n^{(1)}&=&\cG^-_{n-\half}
\,,\EA\EE

\noi
which, after dropping the superscript $^{(1)}$,
reproduces our spin-1 construction in Sect.~3.1, with

\BE\hL_m=L_m+\half Q(m+1)I_m\,.\EE

\noi
Alternatively, to make $\cG^-$ of spin 2, we set
\BE\new\BA{rclcrcl}
\cL^{(2)}_m&=&\multicolumn{5}{l}{\cL_m-\half(m+1)\cH_m\,,}\\
c^{(2)}_m&=&c_{m-\half}\,,&\qquad &b^{(2)}_m&=&b_{m+\half}\,,\\
\cH^{(2)}_m&=&-\cH_m\,,&{}&{}&{}&{}\\
\cG^{(2)}_m&=&\cG_{m+\half}^-\,,&{}&
\cQ_n^{(2)}&=&\cG^+_{n-\half}\,,\\
\hL_m&=&\multicolumn{5}{l}{L_m+\half Q_{{\mathrm{L}}}(m+1)I_m\,,}
\EA\EE

\noi
which leads us back to the above spin-2 construction.

The r\^oles of the BRST operator $\cQ_0$ and its spin-2 superpartner
$\cG_0$ are therefore `dual' to each other in the two twisted
constructions, with one of these operators being `simple'
(coinciding with the $b$ field) and the other one `complicated'. To
preserve the physical content (the cohomologies), one should therefore
consider equivariant cohomologies, thus introducing both
$\cQ_0$ and $\cG_0$ into the consideration, so as to ensure the
equivalence between the two twisted versions.

\def\theequation{\thesection.\arabic{equation}}

\section{Dressed null states via a direct construction}\lvm
On the way from the twisted $N\!=\!2$ symmetry to the \V\ \cs\
imposed by the operators \req{Lontau},
we have
arrived at the `dressed \de s', such as
\req{mastereq} and \req{2}--\req{decouplingop3first}.
These can also be obtained by a direct construction,
which may be useful in practice. The construction amounts to
specifying a way to
restrict from the most general tensor product \de\
in the matter\,$\otimes$\,`Liouville' theory.
As was noted
above, a characteristic feature that the null states \req{UUpsilon} and
\req{UUpsilon3} inherit
from the BRST invariance, is the absence of the {\it
a priori\/} possible ${I_{-1}}^l$
terms where $l$ is the level.
This property, as we will see, can be used to characterize
the required null states, and is one out of the two
basic conditions that we impose, the other one being
the `\K-\M' condition on dressing field operators. Together, these two
will bring
the corresponding decoupling operators to the final form amenable to the
\K-\M\ transform, which at the same time coincides with what
would follow from the $N\!=\!2$ machinery.
We first demonstrate this approach for level 3, and then use it
to arrive at the appropriately dressed \de\ at level 4.
The reader wishing to perform the next step,
the actual transformation to the \V\ \cs, may skip to Sect.~5.

In the tensor product
theory comprising the \emt\ $T(z)=\sum_{n\in {\open
Z}}L_nz^{-n-2}$ and the $U(1)$ current $I(z)=\sum_{n\in{\open Z}}I_n
z^{-n-1}$,
with the commutation relations as written out in \req{thetheory},
let $\Psi$ be a primary field with conformal dimension $\Delta$ and $U(1)$
charge $\hn$.
We now consider levels 3 and 4 separately.

\subsection{Dressing at level 3}\lvm It is
straightforward to check that the most general level-3
null vector w.r.t.\ the semidirect product of \V\ with the $U(1)$
will result from the action on
$|\Psi\rangle$ of the following operator:

\BE\new\BA{rcl} {\cal O}&\equiv&\delta
L_{-3}-2L_{-2}L_{-1}+{1\over\delta+1}L_{-1}^3
-{\delta+1-3\hn^2\over\delta+1}I_{-1}^2L_{-1}
-\hn{2\delta-1\over\delta+1}I_{-2}L_{-1}-2\hn I_{-1}L_{-2}\\
{}&{}&{}+{3\hn\over\delta+1}I_{-1}L_{-1}^2+{\delta^2+\delta-2\delta
\hn^2+\hn^2\over\delta+1}I_{-1}I_{-2}
-\hn{\delta+1-\hn^2\over\delta+1}I_{-1}^3
+\hn{\delta(\delta-1)\over\delta+1}I_{-3}\EA\label{decouplingopfull}\EE

\noi
provided $\delta\equiv\delta_i$, which is the {\sl matter\/} dimension
of $\Psi$, according to

\BE\delta=\Delta-\L(-{\hn^2\over 2}\R)\,,\label{dD}\EE

\noi
is given by

\BE\delta={7-d\mp\sqrt{(1-d)(25-d)}\over 6}\,.\label{delta}\EE

\noi
(It is understood that for the $(p\pr,p)$ minimal model,
$d=1-{6(p\pr-p)^2\over p\pr p}$.)

As explained
above, we can convert the condition $\cO\ket\Psi=0$ into a \de\
of the form of \req{2}, in which
this time we have

\BE\new\BA{rcl}\widehat{\cal O}&\equiv&{1\over\delta+1}{\d^3\over\d z_i^3}
-\delta\sumji\!\L\{{1\over(z_j-z_i)^2}{\d\over\d
z_j}-{2\Delta_j\over(z_j-z_i)^3}\!\R\}+ 2\sumji\!\L\{{1\over
z_j-z_i}\dd{z_j}-{\Delta_j\over(z_j-z_i)^2}\!\R\}\dd{z_i}\\
{}&{}&{}-{\delta-3\hn^2+1\over\delta+1}\!\!
\sum_{{j,~\!k\atop j\neq i,~\!k\neq
i}}\!{n_jn_k\over(z_j-z_i)(z_k-z_i)}\dd{z_i}
+\hn{2\delta-1\over\delta+1}\sumji{n_j\over(z_j-z_i)^2}\dd{z_i}\\
{}&{}&{}-2\hn\!\!\sum_{{j,~\!k\atop j\neq i,~\!k\neq
i}}\!{n_j\over(z_j-z_i)(z_k-z_i)}\dd{z_k}
-{3\hn\over\delta+1}\sumji{n_j\over z_j-z_i}\ddsc{z_i}\\
{}&{}&{}+{\delta^2
+\delta-2\hn^2\delta+\hn^2\over\delta+1}\!\!\sum_{{j,~\!k\atop
j\neq i,~\!k\neq i}}\!\!{n_jn_k\over(z_j-z_i)(z_k-z_i)^2}
+2\hn\!\!\sum_{{j,~\!k\atop j\neq
i,~\!k\neq i}}\!\!{n_j\Delta_k\over(z_j-z_i)(z_k-z_i)^2}
\\
{}&{}&{}+
\hn{\delta+1-\hn^2\over\delta+1}\!\!\!\!\sum_{{j,~\!k,~\!l\atop j\neq
i,~\!k\neq i,~\!l\neq i}}
\!\!\!{n_jn_kn_l\over(z_j-z_i)(z_k-z_i)(z_l-z_i)}-
\hn{\delta(\delta-1)\over\delta+1}\sumji{n_j\over(z_j-z_i)^3}\,.
\EA\label{de3full}\EE

\noi
However, the null vector ${\cal O}\ket\Psi$ with ${\cal O}$ as in
\req{decouplingopfull} allows too much arbitrariness,
as the values of $\Delta$
and $\hn^2$ cannot be fixed separately: there is
a one-parametric freedom which reflects the fact that
we have extended the matter theory by a current.
This extra freedom can be
killed by restricting to a more special form of the operator
\req{decouplingopfull},
achieved by demanding that the coefficient in front
of the $I_{-1}^3$ term vanish, which amounts to setting

\BE\delta=\hn^2-1~.\label{relation}\EE

The next step consists in restricting to a
subsector of those operators whose dimensions and
$U(1)$ charges satisfy the `\K-\M' dressing condition
\req{dressing3}.
This allows us, in the \de, to get rid of the terms

\BE\sum_{{j,~\!k\atop j\neq i,~\!k\neq i}}\!\!
{\hn_j\over (z_i-z_j)(z_k-z_i)^2}
\L\{(2-\hn^2)n_k+2\hn\Delta_k\R\}\biggl\langle\Psi(z_i)\prod_{j\neq
i}\Psi_j(z_j)\biggr\rangle\,,\label{unwanted}\EE

\noi
which do not have analogues in \req{2}, \req{decouplingop3first}.
As a result of \req{relation} and
\req{dressing3},
the decoupling operator \req{de3full}
takes exactly the form  \req{decouplingop3first}.
The present derivation allows us to see by what kind of
restrictions this special form is characterized among
the general family \req{de3full}.
In the next subsection we demonstrate how a
similar argument allows us to arrive at the dressed
null vectors at level 4.

\subsection{Dressing at level 4}\lvm
At level 4, the null vectors exist over `41' (and `14')
and `22' states, and so do of course the dressed null vectors.
We are going to consider them separately.

\subsubsection{Dressed null vectors over the `41' state}\lvm
Consider first the `41' case. The general null vector in the
semidirect product of Virasoro with $U(1)$ reads

\BE\new\BA{l}\ket{\Upsilon_{41}}=\new\BA[t]{l}\L(
25(2\Delta+\hn^2+3)L_{-2} L_{-1}^2+
50(2\Delta+\hn^2+3)\hn I_{-1} L_{-2} L_{-1}\R.\\
{}-(8\hn^4+32\hn^2\Delta+23\hn^2+32\Delta^2+46\Delta-3)L_{-3}L_{-1}
-75\hn I_{-1}L_{-1}^3\\
{}-\frac{75}{4}L_{-1}^4
+25(-2\hn^3+2\hn\Delta+3\hn)I_{-1}^3L_{-1}
+25(\frac{3}{2}-4\hn^2+\Delta)I_{-1}^2L_{-1}^2\\
{}+25(\hn^3+2\hn\Delta-\frac{3}{2}\hn)I_{-2}L_{-1}^2
-(27+3\hn^4+12\hn^2\Delta+18\hn^2+12\Delta^2+36\Delta)
L_{-2}^2\\
{}+(\frac{81}{5}\!
+\frac{8}{5}\hn^6\!+\frac{48}{5}\hn^4\Delta+\frac{32}{5}\hn^4\!
+\frac{96}{5}\hn^2\Delta^2\!
+\frac{128}{5}\hn^2\Delta+\frac{51}{5}\hn^2\!
+\frac{64}{5}\Delta^3\!
+\frac{128}{5}\Delta^2\!+\frac{102}{5}\Delta)L_{-4}\\
{}+(-8\hn^5-32\hn^3\Delta+27\hn^3
-32\hn\Delta^2+54\hn\Delta+3\hn)I_{-3}L_{-1}\\
{}
+(-27+22\hn^4+38\hn^2\Delta+57\hn^2-12\Delta^2-36\Delta)I_{-1}^2L_{-2}\\
{}-(6\hn^5+24\hn^3\Delta+11\hn^3+24\hn\Delta^2
+22\hn\Delta-21\hn)I_{-2}L_{-2}\\
{}-(8\hn^5+32\hn^3\Delta+23\hn^3 +
32\hn\Delta^2+46\hn\Delta-3\hn)I_{-1}L_{-3}\\
{}+
(3+42\hn^4+68\hn^2\Delta-98\hn^2-32\Delta^2-46\Delta)I_{-2}I_{-1}L_{-1}\\
{}+(-\frac{27}{4}-7\hn^4+22\hn^2\Delta+33\hn^2-3\Delta^2-9\Delta)I_{-1}^4
\EA\\
{}+(\frac{8}{5}\hn^7+\frac{48}{5}\hn^5\Delta-\frac{38}{5}\hn^5
+\frac{96}{5}\hn^3\Delta^2-\frac{152}{5}\hn^3\Delta+\frac{6}{5}\hn^3
+\frac{64}{5}\hn\Delta^3-\frac{152}{5}\hn\Delta^2+\frac{12}{5}\hn\Delta
+\frac{27}{10}\hn)I_{-4}\\
{}+(\frac{81}{10}-\frac{11}{5}\hn^6
-\frac{36}{5}\hn^4\Delta+\frac{51}{5}\hn^4
-\frac{12}{5}\hn^2\Delta^2+\frac{134}{5}\hn^2\Delta-\frac{63}{20}\hn^2
+\frac{32}{5}\Delta^3+\frac{64}{5}\Delta^2+\frac{51}{5}\Delta)I_{-2}^2\\
{}+(-\frac{54}{5}-\frac{32}{5}\hn^6
-\frac{112}{5}\hn^4\Delta+\frac{152}{5}\hn^4
-\frac{64}{5}\hn^2\Delta^2+\frac{338}{5}\hn^2\Delta-\frac{24}{5}\hn^2
+\frac{64}{5}\Delta^3+\frac{68}{5}\Delta^2
-\frac{78}{5}\Delta)I_{-3}I_{-1}\\
\phantom{\ket\Upsilon=\L(\R.}
\L.~~{}+(14\hn^5+6\hn^3\Delta-66\hn^3-44\hn\Delta^2-57\hn\Delta
+\frac{27}{2}\hn)I_{-2}I_{-1}^2\R)\ket{\Psi_{41}}\,.\label{41gen}\EA\EE

\noi
Here, as before,
the `dressed' dimension $\Delta$ is related to the matter
dimension $\delta$ of $\Psi$ via eq.\ \req{dD}, while

\BE\delta={41-5d+5\sqrt{(1-d)(25-d)}\over16}\,.\label{delta41}\EE

Now, there are two possibilities to get rid of the $I_{-1}^4$ term.
One of these is to set (cf.\ \req{relation})

\BE\hn^2 = {6\Delta+9\over2}={6\delta+9\over5}\,.\label{4.9}\EE

\noi
Then, eq. \req{41gen} becomes

\BE\ket{\Upsilon_{41}}=\!\!\new\BA[t]{l}
\L(L_{-2}L_{-1}^2+2\hn I_{-1}L_{-2}L_{-1}
-\L({8\hn^2\over15}-1\R)L_{-3}L_{-1}
-{9\over5\hn}I_{-1}L_{-1}^3
-{9\over20\hn^2}L_{-1}^4\R.\\
{}-{4\hn\over5}I_{-1}^3L_{-1}-{11\over5}I_{-1}^2L_{-1}^2
+\L(\hn-{27\over10\hn}\R)I_{-2}L_{-1}^2
-{\hn^2\over5}L_{-2}^2\\
{}+\L({8\hn^4\over45}-{8\hn^2\over15}+{3\over5}\R)L_{-4}
-\L({8\hn^3\over15}-3\hn+{18\over5\hn}\R)I_{-3}L_{-1}
+{4\over5}\hn^2I_{-1}^2L_{-2}\\
{}-\L({2\hn^3\over5}-\hn\R)I_{-2}L_{-2}
-\L({8\hn^3\over15}-\hn\R)I_{-1}L_{-3}
+\L({22\hn^2\over15}-{22\over5}\R)I_{-2}I_{-1}L_{-1}\\
{}+{3\over5\hn}\L({8\hn^6\over27}
-{22\hn^4\over9}+6\hn^2-{9\over2}\R)I_{-4}
{}-{3\over5}\L({5\hn^4\over27}-{11\hn^2\over9}-{33\over4}\R)I_{-2}^2\\
{}-{3\over5}\L({16\hn^4\over27}-{34\hn^2\over9}-27\R)I_{-3}I_{-1}
\L.+\L({4\hn^3\over15}-{6\hn\over5}\R)I_{-2}I_{-1}^2
\R)\ket{\Psi_{41}}\,.\EA\label{null41}\EE

\noi
The corresponding decoupling operator
$\widehat{\cal O}\equiv\widehat{\cO}^{(4)}$
[see \req{2}]
takes the form

\BE\new\BA{l}
\widehat{\cal O}^{(4)}=\!\!\new\BA[t]{l}
{}-{9\over20n_i^2}\d_i^4
+\sumji{1\over z_j-z_i}\L(-\d_j\d_i^2+{9n_j\over5n_i}\d_i^3\R)\\
{}+\sumji{1\over(z_j-z_i)^2}\L\{\!\L({8n_i^2\over15}-1\R)\d_i\d_j
-\L({2n_i\over3}-{6\over5n_i}\R)n_j\d_i^2\R\}\\
{}+\sumji\!\sumki{1\over(z_j-z_i)(z_k-z_i)}\L(
2n_in_k\d_j\d_i-{11\over5}n_jn_k\d_i^2-{n_i^2\over5}\d_k\d_j\R)\\
+\sumji{1\over(z_j-z_i)^3}\L(-{8n_i^3\over45}+{11n_i\over15}-
{3\over5n_i}\R)(n_i\d_j-n_j\d_i)\\
{}+\sumji\!\sum_{k\neq i}{1\over(z_j\!-\!z_i)(z_k\!-\!z_i)^2}\L\{\!
\L(1\!-\!{8n_i^2\over15}
\R)n_j(n_i\d_k-n_k\d_i)+
\L({2\over5}-{4n_i^2\over15}\R)\!n_k(n_i\d_j-n_j\d_i)\R\}\\
{}-\frac{4}{5}\sumji\!\sumki\!\sum_{\l\neq i}
{n_in_kn_l\over(z_j-z_i)(z_k-z_i)(z_l-z_i)}(n_i\d_j-n_j\d_i)\,,
\EA\EA\label{decregular}\EE

\noi
where $\d_j=\d/\d z_j$ (and where, let us recall, $n_i\equiv\hn$).

We have seen that the level-2 and level-3 dressing conditions
\req{dressing2} and \req{dressing3} are described by the
same `universal' (level-independent) formula

\BE
\Delta_j=-\half Qn_j\,,\label{dressinggen}\EE

\noi
where $Q$ is the corresponding
matter \bc, so that the
$\widehat{\phantom{z}}$-dimensions are always zero,
\BE\widehat\Delta_j=0\,,\label{30}\end{equation}
which is what remains of the underlying topological symmetry (zero
`twisted' dimensions of the chiral primary fields).
The same condition \req{dressinggen}
holds in the present case as well,
with the \bc, as deduced from \req{delta41} and \req{4.9},
being given in terms of the
Miwa parameter $\hn\equiv n_i$ as

\BE Q={3\over\hn}-{2\hn\over3}\qquad\qquad (l=4).\label{Jn4}\EE

\noi
In the next section we will show that this is
indeed the \bc\ of the \V\ generators
of the type of \req{Lontau}, through which the operator
$\widehat\cO^{(4)}$
factorizes.

\bigskip

The other possibility to kill the $I_{-1}^4$ term over
the `41' state is

\BE\hn^2={2\Delta+3\over14}={2\delta+3\over15}\,,\label{41coinc}\EE

\noi
with $\delta$ as in \req{delta41}.
The relation between the Miwa parameter
and the \bc\ is now

\BE Q=2\hn-{1\over\hn}\,.\EE

\noi
However, with this value of $Q$ the dressing condition
\req{dressinggen} would not be satisfied unless $\hn^2={1\over4}$.
The matter \cc\ is then $d\!=\!-2$, while
the corresponding null vector
itself is in the intersection of the `22' and `41' null
vectors, which we discuss later, at the end of
Sect.~4.2.2.

\subsubsection{Dressed null vectors over the `22' state}\lvm
As has been noted above, there are more
possibilities at level 4: in addition to \req{41gen},
a null vector exists over the `22' state:

\BE\new\BA{r}\ket{\Upsilon_{22}}
=\new\BA[t]{l}
\L((2\Delta+\hn^2+3)L_{-2}L_{-1}^2+2\hn(2\Delta+\hn^2+3)I_{-1}L_{-2}L_{-1}
+(\hn^2+2\Delta-\frac{3}{2})L_{-3}L_{-1}\R.\\
{}-3\hn I_{-1}L_{-1}^3
-\frac{3}{4}L_{-1}^4
+(-2\hn^3+2\hn\Delta+3\hn)I_{-1}^3L_{-1}
+(\frac{3}{2}-4\hn^2+\Delta)I_{-1}^2L_{-1}^2\\
{}+(\hn^3+2\hn\Delta\!-\!\frac{3}{2}\hn)I_{-2}L_{-1}^2
-(\frac{1}{3}\hn^4+\frac{4}{3}\hn^2\Delta+2\hn^2+\frac{4}{3}\Delta^2
\!+\!4\Delta)L_{-2}^2
+(\frac{3}{2}\hn^2\!+\!3\Delta)L_{-4}\\
{}+(3\hn^3+6\hn\Delta-\frac{3}{2}\hn)I_{-3}L_{-1}
+(\frac{2}{3}\hn^4+\frac{2}{3}\hn^2\Delta+\hn^2-\frac{4}{3}\Delta^2
-4\Delta)I_{-1}^2L_{-2}\\
{}-(\frac{2}{3}\hn^5+\frac{8}{3}\hn^3\Delta+3\hn^3+\frac{8}{3}\hn\Delta^2
+6\hn\Delta-3\hn)I_{-2}L_{-2}\\
{}+(\hn^3+2\hn\Delta-\frac{3}{2}\hn)I_{-1}L_{-3}
+(-\frac{3}{2}+2\hn^4+4\hn^2\Delta-2\hn^2+2\Delta)I_{-2}I_{-1}L_{-1}\\
{}+(-\frac{1}{3}\hn^4+\frac{2}{3}\hn^2\Delta+\hn^2-\frac{1}{3}\Delta^2
-\Delta)I_{-1}^4
+(-\frac{2}{3}\hn^5-\frac{8}{3}\hn^3\Delta+\frac{1}{2}\hn^3
-\frac{8}{3}\hn\Delta^2+\hn\Delta)I_{-4}\\
{}
+(-\frac{1}{3}\hn^6-\frac{4}{3}\hn^4\Delta-\hn^4-\frac{4}{3}\hn^2\Delta^2
-2\hn^2\Delta+\frac{3}{2}\hn^2+\frac{3}{2}\Delta)I_{-2}^2\EA\\
\L.{}+(\frac{8}{3}\hn^4+\frac{14}{3}\hn^2\Delta\!-\!
2\hn^2-\frac{4}{3}\Delta^2\!-\!\Delta)I_{-3}I_{-1}
+(\frac{2}{3}\hn^5+\frac{2}{3}\hn^3\Delta-2\hn^3-\frac{4}{3}\hn\Delta^2
\!-\!\hn\Delta)I_{-2}I_{-1}^2\R)
\!\ket{\Psi_{22}}\EA\label{22}\EE

\noi
As before, we demand that the $I_{-1}^4$ term vanish.
The corresponding equation reads, in
terms of the matter dimension $\delta$ [see \req{dD}],

\BE9\hn^4-12\hn^2\delta-18\hn^2+4\delta^2+12\delta=0,\EE

\noi
whence

\BE\hn^2={2\delta+3\pm3\over3}~.\label{n2delta22}\EE

\noi
If we recall that $\delta$ is now
 related to the matter central charge $d$ via
$\delta={1-d\over8}$, we get

\BE\hn^2={13-d\pm12\over12}~.\label{n2d22}\EE

To see what these two possibilities mean,
consider the
coefficient $C_j$ of the terms

\BE\sum_{j\neq i}\!\sum_{k\neq
i}{C_j\hn_k\over(z_j-z_i)^3(z_k-z_i)}\label{terms}\EE

\noi
in the decoupling equation
corresponding to the null vector \req{22}. It is equal to

\BE C_j=-2\hn
\L(\hn^2+2\Delta-{3\over2}\R)\Delta_j
+\L({8\over3}\hn^4+{14\over3}\hn^2\Delta
-2\hn^2-{4\over3}\Delta^2-\Delta\R)n_j\,,\EE

\noi
where
the notations are the same as before, with $n_j$ and $\Delta_j$ being,
respectively,
the `Liouville' charge and the total dimension of the insertion
$\Psi_j(z_j)$, while $n_i\equiv\hn$ and $\Delta_i\equiv\Delta$.
Similarly to \req{unwanted}, the terms \req{terms} have to vanish
by virtue of a dressing prescription.
The vanishing of $C_i$ gives

\BE\hn^2={2\delta\over3}\qquad{\mathrm{or}}\qquad
\delta={3\over8}\,.\label{extra}\EE

\noi
The first case reproduces the
`$-$' case in \req{n2delta22} and \req{n2d22}.
It will also agree with the dressing condition
\req{dressinggen}, \ie\
$\Delta_j=-\half
Qn_j\equiv-\half\sqrt{(1-d)/3}\,n_j$ for $j=i$ once we choose

\BE\hn=-\sqrt{{1-d\over12}}~.\label{nstrange}\EE

\noi
Then, the corresponding null vector takes the form

\BE\new\BA{rcl}
\ket{\Upsilon_{22}}&=&\L(
\hL_{-2}\hL_{-1}^2+2\hn I_{-1}\hL_{-2}\hL_{-1}
+{\hn^2-\half\over\hn^2+1}\hL_{-3}\hL_{-1}
-{\hn\over\hn^2+1}I_{-1}\hL_{-1}^3\R.\\
{}&{}&{}-{\hn^4+2\hn^2\over\hn^2+1}\hL_{-2}^2
+{{3\over2}\hn^2\over\hn^2+1}\hL_{-4}
+{\hn^3+{\hn\over2}\over\hn^2+1}I_{-3}\hL_{-1}
-{{1\over4}\over\hn^2+1}\hL_{-1}^4\\
{}&{}&{}+{\hn\over\hn^2+1}I_{-1}^3\hL_{-1}
+{\half-\hn^2\over\hn^2+1}I_{-1}^2\hL_{-1}^2
-{{3\over2}\hn\over\hn^2+1}I_{-2}\hL_{-1}^2
-{\hn^2\over\hn^2+1}I_{-1}^2\hL_{-2}\\
{}&{}&\L.{}+\hn I_{-2}\hL_{-2}
+{\hn^3-\half\hn\over\hn^2+1}I_{-1}\hL_{-3}
-{2\hn^2+\half\over\hn^2+1}I_{-2}I_{-1}\hL_{-1}
\R)\ket{\Psi_{22}}\,,\EA\EE

\noi
where $\hL_m$
is defined by the same formula \req{hatshift} as before,
but this time with

\BE Q=-2\hn\qquad\qquad(l=2,\,l'=2).\label{Jn22}\EE

It is now straightforward to
arrive at the \de\ of the type of \req{2} with

\BE\new\BA{c}\widehat{\cal O}^{(22)}
=-\frac{1}{4}\d_i^4
+\sumji{1\over z_j-z_i}\L\{n_in_j\d_i^3
-(n_i^2+1)\d_j\d_i^2\R\}
+\sumji{n_i^3+{n_i\over2}\over(z_j-z_i)^3}(n_i\d_j-n_j\d_i)\\
+\sumji\!\sum_{k\neq i}{1\over(z_j-z_i)(z_k-z_i)}\L\{
(\frac{1}{2}-n_i^2)n_jn_k\d_i^2
+2n_in_j(n_i^2+1)\d_i\d_k
-(n_i^2+2)n_i^2\d_j\d_k\R\}\\
+\sumji{1\over(z_j-z_i)^2}\L\{(\frac{1}{2}-n_i^2)\d_i\d_j
+\frac{3}{2}n_in_j\d_i^2\R\}
+\sumji\!\sum_{k\neq i}\!\sum_{l\neq i}
{n_in_kn_l\over(z_j-z_i)(z_k-z_i)(z_l-z_i)}
(n_i\d_j-n_j\d_i)\\
+\sumji\!\sum_{k\neq i}{1\over(z_j-z_i)(z_k-z_i)^2}
\L\{(n_i^2+1)n_k(n_i\d_j-n_j\d_i)
+(n_i^2-\frac{1}{2})n_j(n_i\d_k-n_k\d_i)\R\}\,.
\EA\label{dec22}\EE

\noi
We proceed with this operator in \req{final22}, where we establish its
relation to the \V\ \cs\ with the \bc\ given by eq.\ \req{Jn22}.

\medskip

The other possibility in \req{n2delta22} is $\hn^2 =
{2\delta\over3}+2={25-d\over12}$.
Therefore, $\Delta=-{d+11\over12}$, which
can be reconciled with the dressing condition
$\Delta=-\half\sqrt{(1-d)/3}\,\hn$
only for the `topological' \cite{[Di]} value $d=-2$ of
the matter central charge [or, $\ctop\!=\!0$ according to \req{d(c)}].
Then,
$\delta={1-(-2)\over8}={3\over8}$, which is
the other case encountered in \req{extra}.
To summarize,

\BE\Delta=-{3\over4}\,,\qquad
Q=1\,,\qquad\hn={3\over2}\,,\qquad d=-2\,,
\label{exceptional}\EE

\noi
which thus applies to a particular type of matter,
that consisting of spin-1 ghosts.
Upon substituting \req{exceptional} into the null vector
\req{22}, we find that the latter becomes identical to the `41'
null vector \req{null41}
with $n_i$ fixed to its value \req{exceptional}, $n_i={3\over2}$.
In fact, the question of when the `41' and `22' null vectors
may coincide can be asked first for the `bare' (purely minimal-model)
null vectors. It is not difficult to see that this does indeed happen
at the above value of the matter \cc\ $d\!=\!-2$ (hence the dimension
$\delta\!=\!{3\over8}$), but also at $d\!=\!28$ ($\delta\!=\!-{27\over8}
$). Then, the corresponding `dressed' null vectors,
\req{41gen} and \req{22}, coincide as well,
and we can ask next if the $I_{-1}^4$
term vanishes. When $d\!=\!-2$, the coefficient in front of
the $I_{-1}^4$ term is proportional to

\BE\L(\hn^2-{9\over4}\R)\L(\hn^2-{1\over4}\R)\,,\label{real}\EE

\noi
while at $d\!=\!28$ this gets replaced by

\BE\L(\hn^2+{9\over4}\R)\L(\hn^2+{1\over4}\R)\,.\EE

\noi
Thus, the `Liouville' charge becomes imaginary for
$d=26-(-2)$. The imaginary unit can be absorbed into
the `liouville' current, thereby
reversing its signature and giving it the r\^ole of a matter,
while the matter itself, by the same procedure, becomes a `Liouville'
theory. Let us therefore concentrate on the case \req{real}, $d\!=\!2$.
Then the \bc\ $Q=\sqrt{{1-d\over3}}=1$, and substituting
this together with $\Delta=\delta-{\hn^2\over2}={3\over8}-{\hn^2\over2}$
into the dressing condition \req{dressinggen}, we find

\BE\hn=\ccases{\frac{3}{2}}{-\frac{1}{2}}\EE

\noi
which belongs to the roots of \req{real}.
The upper value, which we have already met in \req{exceptional},
satisfies also eq.~\req{Jn4}, while the lower one
agrees with both \req{41coinc} and \req{Jn22}. We have thus seen
how the `exceptional' case \req{exceptional}, and the
null vector considered at the end of Sect.~4.2.1, both
fit into the pattern of `intersections' between the `22' and
`41' null vectors.

%
%

\section{From dressed null states
to the Virasoro constraints}\lvm
We have considered in the previous sections the implications
of the `mirror' BRST invariance in the form of the
`\K-\M'
dressed \de s.
In what follows we study its implications
for correlators of ghost-independent chiral primary fields
(hence, as before, of total $\widehat{{\phantom{z}}}$-dimension
zero). We
can observe an interesting property by
invoking the \K-\M\ transform, \ie\ interpreting
the $z_j$ and $n_j$ in \req{decregular}, and \req{dec22}
as the ingredients of the parametrization \req{Miwatransform}.
Namely,
we will show that the decoupling operators
constructed according to the above recipe
`tend to'
factorize through the \V\ generators \req{Lontau}.
The complete factorization occurs for the $(l,1)$
\de s, while for the $(l',l'')$ ones with either
$l'$ or $l''\neq\!1$, there is in general an obstruction to
the factorization. Even in the $(l,1)$ case,
the factorization property is by no
means automatic, and does not apply for instance to the
decoupling operator \req{de3full},
without the relations \req{relation} and \req{dressing3}.

More precisely, let us write the mapping \req{Miwatransform}
as $\cZ\longrightarrow\cT\,:\,(z_j)\mapsto(t_r=
{1\over r}\sumji n_jz_j^{-r})$. Then, at a point $(z_j)$ in the
$\cZ$ space, let $\d/\d z_j$ denote the directions along the fibre
over $(t_r)\in\cT$, so that the general infinitesimal displacements in
$\cZ$ can be written as $\d_j=\dd{z_j}-n_j\sumr z_j^{-r-1}\dd{t_r}$,
where $\dd{z_j}$ does not affect the times. Then it turns
out that, when acting on the function of only
the times $t_r$, the $(l,1)$
decoupling operators always factorize through
the \V\ generators \req{Lontau} in which
the \bc\ $2\sJ-1$ depends on the
level $l$ chosen according to formula \req{Jnl}.

We are going to demonstrate this for the decoupling
operators derived above.

\subsection{Level-2 \de\ as \V\ \cs}\lvm
To transform the decoupling operator from the LHS of \req{mastereq}

\BE\cT=
{1\over 2n_i^2}{\d_i^2}+{1\over n_i}\sum_{j\neq i}{1\over
z_j-z_i}\L(n_i{\d_j}-n_j{\d_i}\R)\label{operator2}
\EE

\noi
into the $t$-variables, we can proceed in a na\"\i ve way by
expressing the derivatives (when these act on
functions depending only on the $t_r$) via

\BE\new\BA{rcl}{\d\over\d z_i}&=&{}
-n_i\sum_{r\geq 1}z^{-r-1}_i {\d\over \d
t_r}~,\\ {\d^2\over\d z_i^2 }
&=&n_i^2\sum_{r,s\geq1}z^{-r-s-2}_i{\d^2\over\d
t_r\d t_s}
+n_i\sum_{r\geq 1}z_i^{-r-2}(r+1){\d\over\d t_r}~,\\ {\d^3\over\d
z_i^3}&=&{}-n_i^3\sum_{q,r,s\geq1}z^{-q-r-s-3}_i{\d^3\over\d t_q\d t_r\d
t_s}-3n_i^2\sum_{r,s\geq1}(r+1)z_i^{-r-s-3}{\d^2\over\d t_r\d t_s}\\
{}&{}&{}-n_i\sum_{r\geq1}(r+1)(r+2)z_i^{-r-3}{\d\over\d t_r}
\EA\label{derivatives}\EE

\noi
(the third derivative is included for future use).
Now, in the second term in $\cT$, we substitute \req{derivatives}
and divide by $z_j-z_i$ to obtain

$$\sum_{j\neq i}n_j\sum_{r\geq1}
\!\sum_{s=1}^{r+1}z_j^{-s}z_i^{-r+s-2}\dd{t_r}~.$$

\noi
Here, the sum over {\it all\/}
$j$ gives $st_s$ according to the \M\ transform
\req{Miwatransform}; the missing term with $j=i$ should be added and
subtracted. Thus

\BE
\sumji{1\over z_j-z_i}(\Rot)=n_i\sum_{r\geq1}
\!\sum_{s=1}^{r+1}z_i^{s-r-2}st_s
\dd{t_r}-n_i^2\sum_{r\geq1}
(r+1)z_i^{-r-2}\dd{t_r}\,,\label{firstorder}\EE

\noi
and therefore, finally,

\BE{\cal T}=\sump z_i^{-p-2}\sL^{(2)}_p\,,\label{3}\EE

\noi
where the $\sL^{(2)}_n$ are given by \req{Lontau} with

\BE2\sJ-1={1\over n_i}-2n_i\qquad\qquad(l=2)\,.\label{Jn2}\EE

\noi
It also follows that the \V\ \bc\ $\sQ\equiv2\sJ-1$ equals
$Q=\sqrt{(1-d)/3}$,
which is the matter \bc.

Therefore, the lowest-level \de\ is nothing but a linear combination
of the $\{p\!\geq\!-1\}$-\V\ generators.
The inverse transformation, {\it from\/} the $\sL_p$ {\it into\/} the
decoupling equation,
was carried out in ref. \cite{[S35]}.
That is, starting with the
`\emt' $\sum_pz^{-p-2}\sL_p$ and evaluating at $z\!=\!z_i$ that part
of it which is holomorphic at the infinity, we find that it equals
the RHS of eq.~\req{operator2}.
The existence of the inverse transform demonstrates that the
(dressed!) correlation functions do indeed depend on the
insertions point only through the time variables $t_r$.

\subsection{\V\ \cs\ from level 3}\lvm
The \de, which is of the third order for level 3, can no longer
just coincide
with the \V\ \cs, which are given by second-order differential operators.
Yet the \V\ \cs\ do emerge anyway.
We will transform
into the time variables
the level-3 decoupling operator \req{decouplingop3first},
$\widehat{\cO}\equiv-n_i\cU$,

\BE\new\BA{rcl}{\cal U}\equiv&{}&{}\\ -{1\over n_i^3}{\d_i^3}
&+&\L(1-{1\over n_i^2}\R)
\sumji{1\over (z_j-z_i)^2} \L(n_i\d_j-n_j
\d_i\R)-{1\over n_i^2}\sumji{1\over z_j-z_i}\L(2n_i{\d_j\d_i
}-3n_j\d_i^2\R)\\ {}&+&{2\over n_i}\sumji\!\sum_{k\neq
i}{n_k\over(z_j-z_i)(z_k-z_i)}\L(n_i\d_j-n_j
\d_i\R)\,,\EA\label{decouplingop3}\EE

\noi
in two ways: the first one is a direct extension of the
above level-2 approach, while the other allows further generalizations,
which, in particular,
will be used at level 4.

\subsubsection{A na\"\i ve approach to level 3}\lvm
Proceeding as in 5.1, by directly dividing over
$z_j-z_i$ as in \req{firstorder},
let us start with the second term
on the RHS of \req{decouplingop3}. We find

\BE\new\BA{l}
{}-\inv{n_i^2}\sumji{1-n_i^2\over (z_j-z_i)^2}\L(n_i\d_j-n_j
\d_i\R)\\
=-\inv{n_i}\sumji{1-n_i^2\over z_j-z_i}n_j\sumr
{z_i^{-r-1}-z_j^{-r-1}\over z_j-z_i}\dd{t_r}\\
{}=\L(n_i-\inv{n_i}\R)\L\{\sumji{n_j\over
z_j-z_i}\!\sumr\!\sum_{s=1}^{r+1}
(z_j^{-s}-z_i^{-s}) z_i^{-r+s-2}\dd{t_r}\!
+\!\sumji{n_j\over z_j-z_i}\sumr z_i^{-r-2}(r+1)\dd{t_r}\R\}\,.\EA
\label{first}\EE

\noi
The first term on the RHS of the last equation,
in its own turn, is equal to

\BE\new\BA{l}
{}-\L(n_i-\inv{n_i}\R)
\sumr\!\sum_{q=1}^{r+1}(r+2-q)z_i^{q-r-3}qt_q\dd{t_r}
-(1-n_i^2)\sumr\!\sum_{q=1}^{r+1}(r+2-q)z_i^{-r-3}\dd{t_r}\EA\EE

\noi
and therefore eq. \req{first} becomes

\BE\new\BA{l}\L(\inv{n_i}-n_i\R)
\sum_{p\geq-1}\!\!\!\sum_{{q\geq1\atop q+p\geq1}}
(p+2)z_i^{-p-3}qt_q\dd{t_{p+q}}\\
{}+(n_i^2-1)\sum_{p\geq1}\frac{1}{2}(p+1)(p+2)z_i^{-p-3}\dd{t_p}
+\L(n_i-\inv{n_i}\R)\sumji{n_j\over z_j-z_i}\!\sum_{p\geq1}z_i^{-p-2}
(p+1)\dd{t_p}~.\EA\EE

Similarly, in the last term in \req{decouplingop3} we can also divide by
$z_j-z_i$ in the following way:

\BE\new\BA{l} {2\over n_i}
\sum_{k\neq i}{n_k\over z_k-z_i}\sumji n_in_j\sumr
{z_i^{-r-1}-z_j^{-r-1}\over z_j-z_i}\dd{t_r}\\
{}
=2\sumji{n_j\over z_j-z_i}\sum_{p\geq-1}\!\!\!
\sum_{{s\geq1\atop p+s\geq1}}
st_sz_i^{-p-2}\dd{t_{p+s}}-2n_i\sumji{n_j\over
z_j-z_i}\sum_{p\geq1}(p+1)z_i^{-p-2}\dd{t_p}~.\EA\EE

Further, the term with second derivatives in \req{decouplingop3}
equals

\BE\new\BA{l}
{}2\sum_{p\geq0}z_i^{-p-3}\sum_{s=-1}^{p-1}\!\!\!
\sum_{{q\geq1\atop q\geq1-s}}
qt_q{\d^2\over\d t_{p-s}\d t_{q+s}}-
2n_i\sum_{p\geq2}z_i^{-p-3}\sum_{s=1}^{p-1}(s+1){\d^2\over\d t_{p-s}\d
t_s}\\
{}-\inv{n_i}\sumji{1\over z_j-z_i}
\L(-n_in_j\sumr\!\sums z_i^{-r-s-2} {\d^2\over\d t_r\d
t_s}-3n_j\sumr(r+1)z_i^{-r-2}\dd{t_r}\R)~.\EA\EE

Substituting also the expression \req{derivatives} for
the third derivative present in the decoupling operator
\req{decouplingop3} and collecting everything together we find

\BE\new\BA{rcl}
{\cal U}&=&\sum_{q,r,s\geq1}z_i^{-q-r-s-3}{\d^3\over\d t_q\d
t_r\d t_s}+\L({3\over
n_i}-2n_i\R)\sum_{r,s\geq1}z_i^{-r-s-3}(r+1){\d^2\over\d t_r\d t_s}\\
{}&+&\L({1\over
n_i^2}+{n_i^2\over2}-\half\R)
\sumr(r+1)(r+2)z_i^{-r-3}\dd{t_r}+{1-n_i^2\over
n_i}\sum_{p\geq-1}\!\!\!\sum_{{q\geq1\atop q\geq1-p}}\!z_i^{-p-3}
(p+2)qt_q\dd{t_{p+q}}\\
{}&+&2\sum_{p\geq0}z_i^{-p-3}\sum_{s=-1}^{p-1}\!\!\sum_{{q\geq1\atop
q\geq1-s}}\!qt_q{\d^2\over\d t_{p-s}\d t_{q+s}}+2\sumji {n_j\over z_j-z_i}
\sum_{p\geq-1}\L(z_i^{-p-2}-z_j^{-p-2}\R)\sL^{(3)}_p\\
{}&{}&\hfill{}+2\sumji{n_j\over z_j-z_i}\sum_{p\geq-1}z_j^{-p-2}
\sL_p^{(3)}\,,
\EA\label{LL}\EE

\noi
where

\BE\new\BA{rcl}{\Sc L}^{(3)}_{p\geq1}&=&\half\sum^{p-1}_{s=1}{\d^2\over\d
t_{p-s}\d t_s}+\sum_{s\geq1}st_s\dd{t_{p+s}}+
{2-n_i^2\over2n_i}(p+1)\dd{t_p}\\ \sL_0^{(3)}&=&\sums st_s\dd{t_s}\\ {\Sc
L}_{-1}^{(3)}&=&\sums(s+1)t_{s+1}\dd{t_s}\EA\label{Lbar}\EE

\noi
are \V\ generators, which differ, however,
from the
ones we have encountered in \req{3}, \req{Jn2}. That is, the
`improvement' term is such that the spin $\sJ$ is related to $n_i$ via

\BE{2\over n_i}-n_i=2\sJ-1\qquad\qquad(l=3)\,.\label{Jn3}\EE

\noi
Note that the last
term has been added to and subtracted from the RHS of \req{LL}. With the
combination
$(z_i^{-p-2}-z_j^{-p-2})/(z_j-z_i)$ we proceed as above, and in
this way get new terms

\BE 2\sum_{p\geq-1}\sum_{s=1}^{p+2}z_i^{-p+s-3}st_s \sL^{(3)}_p
-2n_i\sum_{p\geq-1}z_i^{-p-3}(p+2)\sL^{(3)}_p\,.\EE

\noi
Therefore,

\BE{\cal U}=\sum_{p\geq-2}z_i^{-p-3}\sU_p+2\sumji{n_j\over
z_j-z_i}\sum_{p\geq-1}z_j^{-p-2}\sL^{(3)}_p\,,\label{linearcomb}\EE

\noi
where

\BE\new\BA{rclclcl}\sU_{-2}&=&2\sumr rt_r\sL^{(3)}_{r-2}&{}&{}&{}&{}\\
\sU_{-1}&=&2\sumr rt_r\sL^{(3)}_{r-1}&-&2n_i\sL^{(3)}_{-1}
&+&{1-n_i^2\over
n_i}\sum_{r\geq2}rt_r\dd{t_{r-1}}\\ \sU_0&=&2\sumr rt_r\sL^{(3)}_r&-&4n_i
\sL^{(3)}_0&+&2{1-n_i^2\over n_i}\sumr rt_r\dd{t_r}~~
+~~2\sum_{r\geq2}
rt_r{\d^2\over\d t_1\d t_{r-1}}\\ \sU_{p\geq1}&=&2\sumr rt_r
\sL^{(3)}_{r+p}&-&2(p+2)n_i\sL^{(3)}_p&+&(p+2){1-n_i^2\over n_i}\sumr
rt_r\dd{t_{r+p}}\\
{}&{}&{}&{}&{}&{}&\hfill+~~2\sum_{s=-1}^{p-1}\!\!\sum_{{r\geq1\atop
r\geq1-s}}rt_r{\d^2\over\d t_{p-s}\d t_{r+s}}\\
{}&{}&\multicolumn{5}{l}{+~(p\!+\!1)(p\!+\!2)\!\L({1\over
n_i^2}\!+\!{n_i^2\over2}\!-\!\half\R)\!\dd{t_p}+\L({3\over
n_i}\!-\!2n_i\R)\!\sum_{r=1}^{p-1}(r\!+\!1){\d^2\over\d t_r\d t_{p-r}}
+\!\!\sum_{{q,r,s\geq1\atop q+r+s=p}}\!\!{\d^3\over\d t_t\d t_r\d
t_s}}\EA\label{w3generators}\EE

We can further rewrite the generic $\sU_p$ as

\BE
\sU_p=2\sum_{q=-1}^{p-1}\dd{t_{p-q}}\circ\sL^{(3)}_q+
2\sumr rt_r\sL^{(3)}_{r+p}+
(p+2)\L({1\over n_i}-3n_i\R)\sL_p^{(3)}\,,\qquad p\geq-2\,,\label{final3}
\EE

\noi
which shows that the $\sU_n$ factorize through the \V\ generators, and
thus the $\sU_n$ will
annihilate the tau function once the $\sL^{(3)}_p$ do:
\BE\sL^{(3)}_p\tau=0,\quad
p\geq-1~.\label{L'ontau}\EE

Thus the \V\ generators with the \bc\ tuned according to \req{Jn3}
are hidden in the dressed \de\ (or, in the BRST-invariance
condition in our realization of the topological algebra).
It follows from \req{Jn3} and \req{nsquared3} that
(as was the case at level 2) the minimal matter \bc\ coincides with
the one involved in the \V\ generators,
$\sqrt{(1-d)/3}\equiv Q=2\sJ-1$.

The structure of the
operators $\sU_p$ can be described in slightly more `invariant'
terms, as follows.
Introduce a current $\sI(z)=\sum_{m\in\oZ}z^{-m-1}\sI_m$,
where\footnote{$\sI_0$, which is a c-number, is the
`zeroth time' of the KP \h\ that distinguishes between its different
Schlesinger-transformed
copies \cite{[GO]} (or essentially the parameter $N$
from ref.\ \cite{[S10]}).
However, its value is irrelevant for the following
and can be assumed to be equal to zero.}

\BE\new\BA{rcl} \sI_{m>0}=\dd{t_m}~,\\ \sI_{m<0}=-mt_{-m}~,\EA
\Longrightarrow
\ccases{[\sI_m,\sI_n]=m\Kr{m+n}~,}{[\sL^{(3)}_m,\sI_n]=-n\sI_{m+n}~.}\EE

\noi
Then,

\BE\sU_p=2\sum_{n\leq
p+1}\sI_n\sL^{(3)}_{p-n}+(p+2)\L({1\over n_i}-3n_i\R)\sL_p^{(3)}
\,.\EE

\noi
Obviously, any commutator of
the $\sU$'s
will always factorize in a similar way, with the $\sL$'s on the
right.

As the \cs\ reduce to the \V\ ones, their consistency with the KP flows is
achieved
trivially by virtue of the results of \cite{[S10]}.
Therefore, the initial \de\ also must be consistent with the
Miwa-transformed
version of the KP evolutions.

The generators \req{w3generators} were first derived from the
dressed \de\ in \cite{[GS]}
although they were misinterpreted there.
As we see, the actual mechanism
is that the third-order differential operator factorizes
through the \V\ generators. This property persists at
higher levels, but already at level 4 the factorization
requires a more `invariant' derivation, which will be
first demonstrated below on the already familiar
level-3 example.

\subsubsection{Level-3 factorization once more\label{subsnew}}\lvm
The above derivation of eq.\ \req{linearcomb}, first carried out in
ref.\ \cite{[GS]}, might be
fraught with an ambiguity,
as the times $t_r$ have been viewed in the formulae such as
\req{w3generators} as being
independent of the $z_j$; however, it was not clear from the above
whether indeed
all the $z$'s should be placed
to the left of the $\d/\d t_r$, as was the case in \req{linearcomb}.
At higher levels, a similar ambiguitiy becomes more severe,
so we now present an alternative derivation of \req{linearcomb}, which
avoids the ordering problem (while leading to the same result).

Our starting point will be the previously established
eq.\ \req{3}, which we rewrite in the
form

\BE\new\BA{rcl}\d_i^2&=&{}-2n_i\sumji{1\over z_j-z_i}(\Rot)\\
{}&{}&{}+2n_i^2\sump z_i^{-p-2}\sL^{(3)}_p-(n_i+n_i^3)\sumr z_i^{-r-2}
(r+1)\dd{t_r}\,,\EA\label{d2first}\EE

\noi
where we have anticipated the result by identifying on the RHS the
level-3
\V\ generators \req{Lbar}.

Now, turning to \req{decouplingop3}
we substitute \req{d2first} into
the term $-\inv{n_i^3}\d_i^3=
-\inv{n_i^3}\d_i\circ\d_i^2$ and then again into the resulting
expression $\sumji{n_j\over z_j-z_i}\d_i^2$. We get in this way

\BE\new\BA{rcl}
\cU&=&\d_i\circ{}\L\{-{2\over n_i}\sump z_i^{-p-2}
\sL^{(3)}_p+\L(\inv{n_i^2}+1\R)\sumr z_i^{-r-2}(r+1)\dd{t_r}\R\}\\
{}&{}&{}+2\sumji{n_j\over z_j-z_i}\sump z_i^{-p-2}\sL^{(3)}_p
+\half(n_i^2+1)\sumr(r+1)(r+2)z_i^{-r-3}\dd{t_r}\\
{}&{}&{}-\L(n_i+\inv{n_i}\R)\sumr\!\sum_{s=1}^{r+1}(r+2-s)st_sz_i^{s-r-3}
\dd{t_r}\,.
\EA\EE

\noi
Here, $\d_i$ is a differential operator, composed with the
operator inside the braces, which can be represented as
$\d_i=\dd{z_i}-n_i\sumr z_i^{-r-1}\dd{t_r}$, where
$\dd{z_i}$ does {\it not\/} affect the $t_r$ and acts
 only on the explicit
occurrences
of $z_i$. The operator $\cU$ is understood to act
on functions which depend on only the times $t_r$. Thus,

\BE\d_i\circ\Biggl\{\ldots\Biggr\}=\new\BA[t]{l}
2\sumr\sump z_i^{-r-p-3}\dd{t_r}\circ\sL^{(3)}_p
+\L(\inv{2n_i}-{n_i\over2}\R)\sum_{r,s\geq1}z_i^{-r-s-3}
(r+s+2){\d^2\over\d t_r\d t_s}\\
{}+{2\over n_i}\sumr\!\sum_{s=1}^{r+1}(r+2-s)st_s z_i^{s-r-3}\dd{t_r}
+\L(\inv{n_i^2}-2\R)\sumr(r+1)(r+2)z_i^{-r-3}\dd{t_r}\,.\EA
\label{polefree}\EE

Inserting \req{polefree} into the above expression for
$\cU$, we find

\BE\new\BA{rcl}\cU&=&
2\sumr\sump z_i^{-r-p-3}\dd{t_r}\circ{}\sL^{(3)}_p
+2\sumji{n_j\over z_j-z_i}\sump z_i^{-p-2}\sL^{(3)}_p
\\
{}&{}&{}+\L(\inv{n_i}-n_i\R)\sump(p+2)z_i^{-p-3}\sL^{(3)}_p\,,
\EA\EE

\noi
which is the same as \req{linearcomb}, \req{final3}.
Now we will see how the same approach works at level 4.

\subsection{Factorization at level 4}\lvm
Recall that we had in Sect.~4 more than one relevant null vectors at
level 4.
We will use the method
of section \ref{subsnew} to try to
obtain factorization of the decoupling
operators, after the \K-\M\ transform, into an expression of the form

\BE\widehat\cO=\cA\circ{}\sump z_i^{-p-2}\sL_p^{(*)}\,,\EE

\noi
where $\sL^{(*)}$ are \V\ generators with the expected \bc s:
either \req{Jn4} for the `41' case,
or, for the decoupling
operator \req{dec22}, the \bc\
$\sQ=-2n_i$, as suggested by \req{Jn22}.
The `regular' case, which reproduces the features
we have observed for the `21' and `31' operators,
is provided by the `41' null vector
\req{null41}.

\subsubsection{Virasoro constraints from
the `41' decoupling operator} \lvm
The factorization of the decoupling operator \req{decregular}
will come about as a result of an interplay of the various coefficients,
which appears rather miraculous in the present straightforward
approach,
and takes some work to be established. Our final result for the
\K-\M-transformed decoupling operator \req{decregular}
is given in eq.\ \req{final4}
below. Although a result as simple must have a simpler
derivation, all that we can suggest at the moment
is the following collection of technicalities.

To simplify things, we will utilize in the derivation both the level-2
and level-3 factorization identities found previously.
It is extremely useful to rewrite the level-3 one, eq.\ \req{final3},
as

\BE\new\BA{rcl}\d_i^3&=&{}
{}2n_i^2\nabla_i\circ{}\sV3(z_i)-(n_i^4+n_i^2){\sV3}'(z_i)
+(n_i^3-n_i)\sumji{1\over(z_j-z_i)^2}(\Rot)\\
{}&{}&{}-n_i\sumji{1\over z_j-z_i}(2n_i\d_i\d_j-3n_j\d_i^2)
+2n_i^2\sumji\!\sumki{n_k\over(z_j-z_i)(z_k-z_i)}(\Rot)\,,\EA
\label{d3}\EE

\noi
where we have introduced the compact notations

\BE\sV3(z_i)=\sump z_i^{-p-2}\sL^{(3)}_p\label{vir3}\EE

\noi
and

\BE\sV3'(z_i)=\sump z_i^{-p-3}(-p-2)\sL^{(3)}_p\EE

\noi
for linear combinations of the level-3 \V\ generators \req{Lbar},
while

\BE\nabla_i=\d _i-n_i\sumji{n_j\over z_j-z_i}\EE

\noi
is another very useful combination.

Let us call inv the space of functions which depend on
the $z_j$ only through the times $t_r$. It is understood that
all the operators are considered on this subspace; thus, we are in
fact evaluating $\widehat\cO^{(4)}\L|_{{\mathrm{inv}}}\R.$, which,
however, will not be indicated explicitly.

We substitute \req{d3} into \req{decregular}, expressing
$\d_i^4$ as $\d_i{\circ}{}
\d_i^3$. Then, we get other $\d_i^3$-terms sitting
over a pole in $z_j-z_i$. Substituting \req{d3} again, we then collect
together all the $\d_i^2$-terms and use for these the previous identity
\req{3}, rewritten in the current notations as

\BE\d_i^2=-2n_i\sumji{1\over z_j-z_i}(\Rot)+
2n_i^2\sV2(z_i)\,,\label{d2}\EE

\noi
where $\sV2(z_i)\equiv{\cal T}$
is the same as \req{vir3} for the level-2 \V\ generators
[\ie\ those with the \bc\ \req{Jn2}].
By these manipulations, we bring the decoupling operator \req{decregular}
to the form

\BE\new\BA{rcl}\widehat\cO^{(4)}&=&
\L(-{8n_i^3\over45}-{11n_i\over30}+{3\over10n_i}\R)
\sumji{1\over(z_j-z_i)^3}(\Rot)\\
{}&{}&{}+\L({n_i^2\over12}+{7\over20}\R)\!\!\new\BA[t]{l}
\L\{\sumji{1\over(z_j-z_i)^2}\d_i\d_j\R.\\
{}-\sumji\!\sumki{n_k\over(z_j-z_i)^2(z_k-z_i)}(\Rot)\\
\L.{}+2\sumji\!\sumki{n_k\over(z_j-z_i)(z_k-z_i)^2}(\Rot)\R\}\EA\\
{}&{}&{}+\L(-{13n_i^3\over30}
-{6n_i\over5}\R)\sumji{n_j\over(z_j-z_i)^2}\sV2(z_i)\\
{}&{}&{}-{n_i^2\over5}\sumji{1\over z_j-z_i}\d_j\circ{}\sV2(z_i)
{}+{n_i^2\over10}\sumji\!\sumki{n_jn_k\over(z_j-z_i)(z_k-z_i)}\sV2(z_i)\\
{}&{}&{}
-{9\over10}\nabla_i{\circ}\nabla_i\circ\sV3(z_i)+{9\over20}(n_i^2+1)
\nabla_i\circ{}\sV3'(z_i)\,.\EA\EE

Now, in the terms that do not contain an explicit ${\Sc V}(z_i)$
we use an obvious identity

\BE\new\BA{c}
n_i\L(\sumji{1\over(z_j-z_i)^2}\d_i\d_j-\sumji\!\sumki{n_k\over
(z_j-z_i)^2(z_k-z_i)}(\Rot)\R.\\
\hfill\L.{}+2\sumji\!\sumki{n_k\over(z_j-z_i)(z_k-z_i)^2}(\Rot)\R)\\
{}=\nabla_i\circ\sumji{1\over(z_j-z_i)^2}(\Rot)
+2\sumji{n_j\over(z_j-z_i)^2}\sV2(z_i)
-2\sumji{1\over(z_j-z_i)^3}(\Rot)\,.\EA\EE

\noi
Another trivial but useful rearrangement reads

\BE n_i^2\sumji\!\sumki{n_jn_k\over(z_j-z_i)(z_k-z_i)}
=\nabla_i^2-\d_i^2+2n_i\sumji{n_j\over z_j-z_i}\d_i
+n_i\sumji{n_j\over(z_j-z_i)^2}\,.\EE

\noi
Using these allows us to arrive at

\BE\new\BA{rcl}\widehat\cO^{(4)}&=&{}
\L({4n_i^2\over15}+{2\over5}\R)\L(-{1\over n_i}-{2n_i\over3}\R)
\sumji{1\over(z_j-z_i)^3}(\Rot)\\
{}&{}&{}+\L({n_i\over12}+{7\over20n_i}\R)\nabla_i\circ\!\!
\sumji{1\over(z_j-z_i)^2}(\Rot)\\
{}&{}&{}
+\L(-{4n_i^3\over15}-{2n_i\over5}\R)\sumji{n_j\over(z_j-z_i)^2}\sV2(z_i)
-{n_i\over5}\sumji{1\over z_j-z_i}(\Rot)\circ\sV2(z_i)\\
{}&{}&
{}+{9\over20}(n_i^2+1)\nabla_i\circ{}\sV3'(z_i)-{1\over10}\d_i^2
\circ{}\sV2(z_i)\\
{}&{}&{}
+\L({n_i\over12}+{7\over20n_i}\R)\nabla_i^2
\circ\sumr z_i^{-r-2}(r+1)\dd{t_r}
-{4\over5}\nabla_i^2{\circ}\sV4(z_i)\,.\EA\label{work}\EE

\noi
The last two terms here are just a rewriting of

\BE{1\over10}\nabla_i^2{\circ}
\sV2(z_i)-{9\over10}\nabla_i^2{\circ}\sV3(z_i)
\,,\EE

\noi
and we have introduced
the `level-4' \V\ generators
as suggested by formula \req{Jn4},

\BE\sL_{r\geq1}^{(4)}=\half\sum_{s=1}^{r-1}{\d^2\over\d t_s\d t_{r-s}}
+\sum_{s\geq1}st_s\dd{t_{s+r}}+\L({3\over2n_i}-{n_i\over3}\R)
(r+1)\dd{t_r}\label{L4}\EE

We thus get in \req{work}
two terms with the same coefficient
$\L({n_i\over12}+{7\over20n_i}\R)$, and to
add these together we use the identity

\BE\new\BA{l}
\sumji{1\over(z_j-z_i)^2}(\Rot)+\nabla_i\circ
\sumr z_i^{-r-2}(r+1)\dd{t_r}\\
{}=n_i\sV3'(z_i)\\
{}=n_i\sV4'(z_i)+\L({n_i^2\over6}
+\half\R)\sumr(r+1)(r+2)z_i^{-r-3}\dd{t_r}\EA\label{secondorder}\EE

\noi
Similarly, in the first term in \req{work} we use the identity

\BE\new\BA{rcl}\sumji{1\over(z_j-z_i)^3}(\Rot)&=&
n_i\sumji{n_j\over(z_j-z_i)^2}\sumr(r+1)z_i^{-r-2}\dd{t_r}
+{n_i\over2}\sV4''(z_i)\\
{}&{}&
{}+\half\nabla_i\circ\sumr(r+1)(r+2)z_i^{-r-3}\dd{t_r}\\
{}&{}&{}-{n_i\over2}
\sum_{r,s\geq1}(r+1)(s+1)z_i^{-r-s-4}{\d^2\over\d t_r\d t_s}\\
{}&{}&{}-{1\over4}\sumr(r+1)(r+2)(r+3)z_i^{-r-4}\dd{t_r}\,,\EA\EE

\noi
where

\BE\sV4''(z_i)=\sump\!z_i^{-p-4}(p+2)(p+3)\sL^{(4)}_p\,.\EE

\noi
As a result, the operator $\widehat\cO^{(4)}$ becomes

\BE\new\BA{rcl}\widehat\cO^{(4)}&=&
\L({n_i^2\over12}+{7\over20}\R)\nabla_i{\circ}\sV4'(z_i)
+\L(-{4n_i^3\over15}-{2n_i\over5}\R)\sumji{n_j\over(z_j-z_i)^2}
\sV4(z_i)\\
{}&{}&{}+{9\over20}(n_i^2+1)\nabla_i{\circ}\sV3'(z_i)
-{4\over5}\nabla_i^2{\circ}\sV4(z_i)\\
{}&{}&{}-{n_i\over5}\sumji{1\over z_j-z_i}(\Rot)\circ\sV2(z_i)
-{1\over10}\d_i^2{\circ}\sV2(z_i)\\
{}&{}&{}+{n_i\over5}\L({1\over n_i}+{2n_i\over3}\R)^2\Biggl\{\!
\!\!\new\BA[t]{l}
{}-n_i\sV4''(z_i)
+n_i\sum_{r,s\geq1}(r+1)(s+1)z_i^{-r-s-4}{\d^2\over\d t_r\d t_s}\\
{}+\half\sumr(r+1)(r+2)(r+3)z_i^{-r-4}\dd{t_r}\Biggr\}\EA\\
{}&{}&{}+\L(-{3n_i^3\over40}-{n_i\over6}-{1\over40n_i}\R)\nabla_i\circ
\sumr(r+1)(r+2)z_i^{-r-3}\dd{t_r}\,.\EA\label{work2}\EE

Now, in the term
${9\over20}(n_i^2+1)\nabla_i{\circ}\sV3'(z_i)$ we replace
$\sL^{(3)}_p$
with
$\sL^{(4)}_p$; the difference then adds up with the last term in
\req{work2} to produce the contribution

\BE\L({2n_i\over15}+{1\over5n_i}\R)\nabla_i\circ
\sumr(r+1)(r+2)z_i^{-r-3}\dd{t_r}\,.\label{term}\EE

\noi
On the other hand, we have in \req{work2}

\BE\new\BA{rcl}{}-{n_i\over5}\sumji{1\over z_j-z_i}(\Rot)\circ\sV2(z_i)
{}&=&{}-{n_i\over5}\sump z_i^{-p-2}
\sumji{1\over z_j-z_i}(\Rot)\circ\sL^{(2)}_p\\
{}&{}&{}+{n_i\over5}\sumji{n_j\over z_j-z_i}\sV2'(z_i)\,.\EA\EE

\noi
The last piece here combines with that part of \req{term}
that sits over the pole coming from the definition of
$\nabla_i$, to produce precisely a
$\sV4'(z_i)$ over the pole. Thus,

\BE\new\BA{rcl}\widehat\cO^{(4)}&=&
{4\over5}\L({2n_i^2\over3}+1\R)\nabla_i{\circ}\sV4'(z_i)
{}+\L(-{4n_i^3\over15}-{2n_i\over5}\R)\sumji{n_j\over(z_j-z_i)^2}
\sV4(z_i)\\
{}&{}&{}-{4\over5}\nabla_i^2\circ{}\sV4(z_i)
-{n_i^2\over5}\L({1\over n_i}+{2n_i\over3}\R)^{\!2}\sV4''(z_i)
+{n_i\over5}\sumji{n_j\over z_j-z_i}\sV4'(z_i)\\
{}&{}&{}-{1\over10}\d_i^2\circ{}\sV2(z_i)+
{n_i^2\over5}\L({1\over n_i}+{2n_i\over3}\R)^{\!2}\!
\sum_{r,s\geq1}(r+1)(s+1)z_i^{-r-s-4}{\d^2\over\d t_r\d t_s}\\
{}&{}&{}+{n_i\over10}\L({1\over n_i}+{2n_i\over3}\R)^{\!2}
\sumr(r+1)(r+2)(r+3)z_i^{-r-4}\dd{t_r}\\
{}&{}&{}+{1\over5}\!\L({1\over n_i}+{2n_i\over3}\!\R)\d_i\circ\!
\sumr(r+1)(r+2)z_i^{-r-3}\dd{t_r}\\
{}&{}&{}-{n_i\over5}\!
\sump\!z_i^{-p-2}\sumji\!{1\over z_j-z_i}(\Rot)\circ\sL^{(2)}_p\,.
\EA\label{work3}\EE

\noi
In the last term, there is clearly no pole, by virtue of
\req{firstorder}.

Now, we replace the remaining $\sL^{(2)}$ with
$\sL^{(4)}$, and add the difference
to that part of \req{work3} that is not factorized through the
$\sL^{(4)}$ yet. Rather surprisingly, this gives just

\BE
{n_i\over5}\L({2n_i^2\over3}+1\R)\sumr(r+1)z_i^{-r-2}
\!\sump\!z_i^{-p-2}\dd{t_r}\circ\sL^{(4)}_p\,.\EE

Thus all the terms in the decoupling operator have been factorized
through
$\sL^{(4)}_p$ on the right:

\BE\new\BA{rcl}\widehat\cO^{(4)}&=&
{4\over5}\L({2n_i^2\over3}+1\R)\nabla_i\circ{}\sV4'(z_i)
{}-{2n_i\over5}\L({2n_i^2\over3}+1\R)\sumji{n_j\over(z_j-z_i)^2}
\sV4(z_i)\\
{}&{}&{}-{4\over5}\nabla_i^2\circ{}\sV4(z_i)
-{1\over5}\L({2n_i^2\over3}+1\R)^{\!2}\sV4''(z_i)
-{1\over10}\d_i^2\circ{}\sV4(z_i)\\
{}&{}&{}-{n_i\over5}\sump\!\!z_i^{-p-2}\sumji\!{1\over z_j\!-\!z_i}
(n_i\d_j-n_j\d_i)\circ\sL^{(4)}_p\\
{}&{}&{}+{n_i\over5}\L({2n_i^2\over3}+1\R)\!\sumr(r+1)z_i^{-r-2}\!
\sump\!z_i^{-p-2}\dd{t_r}\circ\sL^{(4)}_p.
\EA\label{final4}\EE

\noi
Along with eqs.\ \req{Jn2} and \req{Jn3},
the \bc\ of the \V\ generators $\sL^{(4)}_p$
follows the general pattern \req{Jnl} that was
conjectured in \cite{[GS]}.

Finally, since we are in fact considering
$\widehat\cO^{(4)}\L|_{{\mathrm{inv}}}\R.$,
the last formula can be rewritten as

\BE\widehat\cO^{(4)}=\cA\circ\!\sump\!z_i^{-p-2}\sL_p^{(4)}\,,
\label{OAL}\EE

\noi
with
\addtocounter{equation}{-2}
\def\theequation{\thesection.\arabic{equation}$'$}

\BE\new\BA{rcl}
\cA&=&\L({2n_i\over5}-{4n_i^3\over15}\R)\sumji{n_j\over(z_j-z_i)^2}
-{4n_i^2\over5}\sumji\!\sumki{n_jn_k\over(z_j-z_i)(z_k-z_i)}\\
{}&{}&{}+
\L({4n_i\over5}-{8n_i^3\over15}\R)\sumji{n_j\over z_j-z_i}\dd{z_i}
-{8n_i^2\over5}\sumji{n_j\over z_j-z_i}\sumr z_i^{-r-1}\dd{t_r}\\
{}&{}&{}-{9n_i^2\over10}\sum_{r,s\geq1}z_i^{-r-s-2}{\d^2\over
\d t_r\d t_s}-{n_i^2\over5}\sumr\!
\sum_{s=1}^{r+1}z_i^{s-r-2}st_s\dd{t_r}\\
{}&{}&{}+\L({n_i^3\over3}-{7n_i\over10}\R)\sumr(r+1)z_i^{-r-2}\dd{t_r}
+\L(n_i-{8n_i^3\over15}\R)\sumr z_i^{-r-1}\dd{t_r}\dd{z_i}\\
{}&{}&{}
+\L(-{4n_i^4\over45}+{4n_i^2\over15}-{3\over10}
\R)\ddsc{z_i}\,,
\EA\EE
\addtocounter{equation}{1}
\def\theequation{\thesection.\arabic{equation}}

\noi
where $\d/\d z_i$ acts only on the explicit occurrences of $z_i$
in \req{OAL}, while the $\d/\d t_r$ are {\it composed\/} with the
$\sL_p^{(4)}$.

Equations \req{final4} constitute our result
for the factorization of the `41' decoupling operator
through the \V\ generators. The pattern of the factorizations
observed so far appears very convincing, and in Sect.~6 we suggest
how the particular cases considered above fit into the
general picture.

\subsubsection{The `22' decoupling operator}\lvm
A somewhat different
situation occurs for the `22' case. One could expect a factorization
of the decoupling operator through the \V\ generators

\BE\sL_{r\geq1}^{(22)}=\half\sum_{s=1}^{r-1}{\d^2\over\d t_s\d t_{r-s}}
+\sum_{s\geq1}st_s\dd{t_{s+r}}-n_i(r+1)\dd{t_r}\,,\label{L22}\EE

\noi
with the \bc\ taken from \req{Jn22}. However,
applying the same strategy
as above to the operator \req{dec22}, we arrive at

\BE\new\BA{l}\widehat{\cal O}^{(22)}=\!\new\BA[t]{l}
\half\nabla_i\circ\sV{22}'(z_i)+\nabla_i^2\circ\sV{22}(z_i)\\
{}+(n_i^2+2)\!\!\new\BA[t]{l}\L\{
-\half\d_i^2\circ
\sV{22}(z_i)-n_i\!\sumji\!{1\over z_j-z_i}(\Rot)\circ\sV{22}(z_i)
-{1\over4}\sV{22}''(z_i)\R.\\
{}\L({n_i^3\over2}\!+\!{1\over2n_i}\R)\!
\sumji\!{n_j\over(z_j\!-\!z_i)^2}\sV{22}(z_i)
\!-{n_i\over2}\sumr(r\!+\!1)z_i^{-r-2}\!\sump\!\!z_i^{-p-2}\dd{t_r}
\circ\sL_p^{(22)}\\
\L.{}+{1\over4}\L(n_i-\inv{n_i}\R)^{\!2}\sumji{n_j\over(z_j-z_i)^2}
\sumr(r+1)z_i^{-r-2}\dd{t_r}\R\}\EA\EA\EA
\label{final22}\EE

\noi
and thus, although many cancellations (which have yet to
be explained in an `invariant' way\footnote{
Despite there being no complete factorization,
the occurrence of the \V\ generators with exactly the `predicted'
\bc\ \req{Jn22} is quite remarkable.})
have occurred en route from
\req{dec22} to \req{final22}, the last term represents
an obstruction to factorization.
Its absence in the $(l,1)$ case might be attributed to the
special features enjoyed by the
$(1,l)$ and $(l,1)$ null vectors \cite{[BSa],[BdFIZ],[FGPP]}.
In Sect.~7 we comment briefly on why the $(l,1)$
case is preferred by the \K-\M\ transform.

The obstruction would only vanish for a special value of
$n_i$ (while in the $(l,1)$
cases considered above, $n_i$ could
be chosen as a free parameter). In view of eq.\ \req{nstrange},
this is $n_i=-1$, and therefore

\BE Q=2,\qquad d=-11\,.\label{211}\EE

\noi
Thus, finally, in this particular case we have the factorization

\BE\new\BA{l}\widehat{\cal O}^{(22)}=\!\new\BA[t]{l}\frac{1}{2}
\nabla_i\circ\sV{\bar{22}}'(z_i)+\nabla_i^2\circ\sV{\bar{22}}(z_i)\\
{}-\frac{3}{2}\d_i^2\circ\sV{\bar{22}}(z_i)
+3\sumji\!{1\over z_j-z_i}(\frac{3}{2}\d_j-n_j\d_i)\circ\sV{\bar{22}}
(z_i)
-\frac{3}{4}\sV{\bar{22}}''(z_i)\\
{}-3\sumji{n_j\over(z_j-z_i)^2}\sV{\bar{22}}(z_i)
+\frac{3}{2}\sumr(r+1)z_i^{-r-2}\!\sump\!z_i^{-p-2}\dd{t_r}
\circ\sL_p^{(\bar{22})}\EA\EA
\label{final22bar}\EE

\noi
with

\BE\sL_{r\geq1}^{(\bar{22})}=
\half\sum_{s=1}^{r-1}{\d^2\over\d t_s\d t_{r-s}}
+\sum_{s\geq1}st_s\dd{t_{s+r}}+(r+1)\dd{t_r}\,.\label{L22bar}\EE

\noi
The numbers \req{211} are characteristic of the spin-${3\over2}$ ghosts
\cite{[FMS]}.
Thus, along with the two ghost systems participating
in the diagram \req{diagram}, another one is `preferred' by the
\K-\M\ transform.

Recall finally that the
$d=-2$ matter corresponds to the set \req{exceptional}.
\K-\M\ transforming the corresponding
`exceptional' decoupling operator
does not require any additional calculations in view
of the coincidences stated at the end of Sect.~4.2.2.

\section{Summary and generalizations}\lvm
The above can be summarized and generalized
as follows. Let us
{\it start\/} from the \V\ \cs\ and ask ourselves
whether they admit
a transformation to the $z_j$ variables introduced according to
\req{Miwatransform}.
The answer
depends on an interplay between the involved parameters,
\ie\ the `spin' $\sJ$ entering the \cs, the Miwa parameter $n_i$,
and an integer $l$, which is to become the level.
For $l\!=\!2$ everything is very simple and the `half' of
the \emt\ itself, $\sT_{\geq-1}(z_i)=\sump z_i^{-p-2}\sL_p$, undergoes
a transformation to the $z_j$ variables. However, for $l\!\geq\!3$
the situation is different, as the \V\
generators by themselves no longer allow a transformation
into the $z_j$ variables,
but only in higher-order
combinations such as \req{linearcomb} or \req{final4},
however unnatural these may appear from the point of view of
the $t$ variables. Fixing the coefficients appropriately and
inverting the previous steps\footnote{Technically, this can be done by
evaluating contour integrals, as in \cite{[S35]}.}, we arrive at the
corresponding $(l,1)$ \de\
in a dressed $(p\pr,p)$ minimal model,
where

\BE{p\pr\over p}=1+{\sQ^2\over4}
\pm{\sQ\over4}\sqrt{\sQ^2+8},\qquad \sQ\equiv2\sJ-1\,,
\label{6.1}\EE

\noi
provided

\BE 2\sJ-1\equiv\sQ=
{l-1\over\hn}-{2\hn\over l-1},\qquad(l\geq2)\,.\label{Deltajl}
\EE

\noi
The integer $l$ specifies the $(l,1)$ \de\ in the dressed
minimal model. Equation \req{6.1} implies
${\displaystyle (2\sJ-1)^2=2{(p-p\pr)^2\over
pp\pr}}$, which, however, is {\it not\/} a mere rewriting of the
standard minimal model formula \req{Q(d)}, since
{\it a priori\/} $\sQ=2\sJ-1$,
as read off from the \V\ \cs, need not coincide with
the minimal model \bc\ $Q=\sqrt{{1-d\over3}}$
(but it does as a result
of the calculation).

To establish the correspondence starting from the minimal model's end,
one first
dresses the minimal model according to the `\K-\M' dressing
prescription which ensures
that all operators have total dimensions proportional to their
`Liouville' charges (to be identified with the Miwa parameters):

\BE\Delta_j=-\half Qn_j\,.\label{gendec}\EE

\noi
The {\it matter\/} dimensions are given by subtracting away the
$U(1)$ Sugawara part, $\delta_j=\Delta_j-\L(-\half n_j^2\R)$, hence
$n_j$ must be determined from the equation
[which we have already met in \req{preliminary}]

\BE n_j^2-Qn_j-2\delta_j=0~.\label{nequation}\EE

\noi
One also imposes the vanishing of the coefficient in front of the
$I_{-1}^l$ term in the null vector build upon
the $(l,1)$ primary state.
Then the order-$l$ \de\ corresponding to a given
insertion of $\Psi\equiv\Psi_{l1}$ at the point $z_i$,

\BE\L\{\d_i^l+\sumji{1\over z_j-z_i}(-ln_in_j\d_i^{l-1}+b_j\d_i^{l-2}\d_j
+\ldots ) +\ldots\R\}\biggl\langle\Psi(z_i)\prod_{j\neq i}
\Psi_j(z_j)\biggr\rangle=0\,,\label{de}\EE

\noi
takes the form, after the \K-\M\ transform, of the constraint

\BE\L\{\sum_{n\geq-l+1}\!z_i^{-n-l}\sU_n^{(l)}+\sumji{k_j\over
z_j-z_i}\sum_{n\geq-l+2}\!\!z_j^{-n-l+1}\sU_n^{(l-1)}+\ldots\R\}\tau(t)=0
\,,\label{constraintl}\EE

\noi
where the omitted terms contain multiple poles, up to the last
group of terms of
order $(l-2)$,
which enter multiplied with $\sU_n^{(2)}\equiv\sL_n$. All
the $\sU$ operators
factorize through the $\sL_n$ on the right.
These \V\ generators are of the form of \req{Lontau} with
$\sJ$ fixed by the minimal model chosen,
according to \req{6.1}.

This suggests the correspondence between
conformal field-theoretic ingredients and
the KP tau function $\tau(t)$
to be achieved via the ansatz \req{firstansatz}. More precisely,
consider, in the dressed minimal model, correlators
involving different
primary fields\footnote{The appropriate
insertions of the \bc\ and/or integrals
of top forms of `topological multiplets'
are understood.},

\BE\tau(t)\equiv\tau\{z_j\}=
\biggl\langle\prod_a
\Psi_{l_a1}(z_{i_a})\biggr\rangle\,,\label{ansatzl} \EE

\noi
where each of the
$\Psi_{l_a1}$ operators has an
$(l_a,1)$ null descendant, the `matter' dimensions of the
$\Psi_{l_a1}$ being given by the RHS of \req{deltagen}.
Then, eq.\ \req{nequation} (in which $Q=\sqrt{{1-d\over3}}$ is fixed)
for $j=i_a$ gives the respective
`Liouville' charges $n_{i_a}$ of the $\Psi_{l_a1}$:

\BE n_{i_a}^2=(l_a-1)^2\,{13-d\pm\sqrt{(1-d)(25-d)}\over24}
={(l_a-1)^2\over2}\L({p'\over p}\R)^{\pm1}.\EE

\noi
Now, the null states built
over the $\ket{\Psi_{l_a1}}$
give rise to the \de s of the form of
\req{de}, the corresponding decoupling operator being
$\d_{i_a}^{l_a}+\ldots$. The possibility to
\K-\M-transform each one of these \de s
depends only on the value of the respective
parameter $n_{i_a}$. However, as we have just seen,
the $n_{i_a}$ are such that
{\it each\/} of them corresponds to the same \bc\

\BE 2\sJ-1=
{l_a-1\over n_{i_a}}-{2n_{i_a}\over l_a-1}=\sqrt{2}{|p'-p|\over\sqrt{p'p}
}\EE

\noi
and therefore
all the corresponding decoupling operators factorize as

\BE\widehat\cO^{(l_a)}_{i_a}
=\cA_a\circ\!\sump\!z_{i_a}^{-p-2}\sL_p\,,\label{ola}\EE

\noi
with {\it the same\/} \V\ generators $\sL_p$ for every $a$.
Thus, a given set of \V\ \cs\ implies all the $(l,1)$ \de s
in the appropriate minimal model. Vice versa, it seems
plausible that the full set of the
$\widehat\cO^{(l_a)}_{i_a}$-\de s will imply the \cs\
$\sump\!z_{i_a}^{-p-2}\sL_p\tau=0$ for every $a$, and these in turn
would allow us to conclude that
$\sL_p\tau=0$, $p\!\geq\!-1$~\footnote{
We are indebted to E.~Kiritsis for a stimulating
discussion of this point.}.

One may also view the dressing prescription \req{gendec} as a
manifestation of the underlying BRST invariance in the
realization of the topological algebra constructed in Sect.~3.1.
The relation between the topological $U(1)$ charge of
ghost-independent
chiral primary states and
the \tcc,

\BE\htop^2-(l-1){\ctop+3\over6}\htop+(l-1)^2\,{\ctop-3\over6}=0\,,
\label{htopctop}\EE

\noi
gives

\BE n_i^2=\ccases{(l-1)^2\,{3-\ctop\over12}}{(l-1)^2\,{3\over3-\ctop}}\EE

\noi
Now, the equations \req{gendec} and
\req{Deltajl} give, for $\Delta\equiv\Delta_i$, which is
the matter+`Liouville'
dimension of $\Psi$:

\BE\Delta={n_i^2\over l-1}+{1-l\over2}\,,\EE

\noi
whence the `minimal' dimension $\delta\equiv\delta_i$
is given by eq.\ \req{deltagen}.

\section{Concluding remarks}\lvm
Using the \K-\M\ transform,
we have related the \V\ \cs\ on the KP \h\
to the highest-weight conditions (including the
BRST invariance)
in a realization of
the topological (twisted $N\!=\!2$)
algebra.
In the representation, constructed in Sect.~3.1, of the
topological algebra \req{topalgebra} in terms of matter,
`Liouville' and $c\!=\!-2$ ghost fields, the BRST invariance can be
imposed level by level, by factoring away BRST-exact highest-weight
vectors. In general, this gives an infinite set of
equations on the correlators of various fields. The analysis carried out
in this paper applies to the case when the correlators contain
ghost-independent representatives of chiral
primary fields (note the BRST operator \req{Q})
and the equations become the appropriately
dressed $(l,1)$ \de s. These latter then factorize through the \V\
generators that constrain the KP tau function.
Unfortunately, we only observe this fact
as an `experimental evidence', and
are unaware of its `invariant' explanation.

It might be worth emphasizing that, although it is the level-2
decoupling operator \req{operator2}
that essentially coincides with the \V\ \cs,
the higher-level
decoupling operators do not factorize through the level-2 one
(which would have been a contradiction); instead,
when {\it keeping $n_i\!=\!\hn$ fixed\/}, so as to have `the same'
\K-\M\ transform for every level,
each of the $(l,1)$ decoupling operators factorizes through a distinct,
$l$-dependent, set of \V\ generators $\sL_{p\geq-1}^{(l)}$, whose \bc\
$\sQ$ carries an explicit dependence on $ l$
as given by \req{Deltajl}.
It is instructive, however, that technically, in order to
prove the factorization, we were
trying to factorize a given decoupling operator
of a given level through the lower decoupling operators
`as far as possible'. The
`leftover', which accounts for the fact that the decoupling operator
is {\it not\/} in the ideal generated
by the lower ones, is such that, {\em after the \K-\M\ transform\/},
it serves precisely to `correct' the \bc s of the \V\ generators
involved.

To return to the interpretation in terms of differential operators, what
we have observed can be stated as a theorem on `meromorphic'
differential operators of order $l$, of the form\footnote{
To ensure that this corresponds to an $(l,1)$ null state,
one sets $a_j=-ln_in_j$ with $n_i=(l-1)\sqrt{{p'\over2p}}$.}

\BE\widehat\cO={\d^l_i}
+\sumji{1\over z_j-z_i}\L(a_j{\d^{l-1}_i}+
b_j{\d^{l-2}_i}{\d_j}+\ldots\R)+\ldots\EE

\noi
with vanishing zeroth-order part, in (infinitely many) variables $z_j$.
Introducing parameters $t_r$ by eq.\ \req{Miwatransform}
and calling inv the subspace of those functions that depend on the $z_j$
only through the $t_r$, we have that
the conditions
\def\hoL{\!\!\hat{\phantom{\cI}\!\oL}}

\BE\new\BA{r}
{[}\widehat{\cO},{\hoL}_1{]}\equiv0{{\mathrm{mod}}}
(\hoL_{\geq1},{\hat{\oI}}_{\geq1})\,,\\
{[}\widehat{\cO},{\hoL}_2{]}\equiv0{{\mathrm{mod}}}
({\hoL}_{\geq1},{\hat{\oI}}_{\geq1})\,,\\
{[}\widehat{\cO},{\hat{\oI}}_1{]}\equiv0{{\mathrm{mod}}}
({\hoL}_{\geq1},{\hat{\oI}}_{\geq1})\,,\EA\EE

\noi
with

\BE\new\BA{rcl}
\hoL_1&=&\sum_j(z_j-z_i)^2\dd{z_j}\,,\\
\hoL_2&=&\sum_j(z_j-z_i)^3\dd{z_j}\,,\\
\hat{\oI}_1&=&\sum_j n_j(z_j-z_i)\,,\EA\EE

\noi
imply that $\widehat\cO\L|_{{\mathrm{inv}}}\R.$ factorizes
through the \V\ generators \req{Lontau}.
This reformulation may provide an appropriate setting for proving the
factorization in general.

An interesting possibility would be
to study the implications for the conformal models of the
Miwa-transformed \cite{[Mi],[Sa]} Hirota bilinear relations.
However, to prove that certain properties (e.\,g.
fusion rules) of
conformal models are equivalent (?) to Hirota-like identities, one has to
take the latter
in the version that, unlike the usual identities, involves
Miwa parameters
shifted as $n_k\mapsto n_k\pm n_j$ rather than $n_k\mapsto
n_k\pm1$. The demonstration of such `discrete' identities would provide
an independent and
rigorous proof for our ansatz \req{ansatzl} for the tau function.

\bigskip

We have also seen that two different
``bosonizations" of the type \hfill\break
\centerline{twisted $N\!=\!2$
$\Longrightarrow$ matter + `Liouville'
+ ghosts}\newline
are possible, resulting in different
prescriptions to dress the
matter. While the more standard DDK case is reproduced by
taking spin-2 ghosts,
it is the spin-1 ghost system that
is required to prove the direct
correspondence with the integrable formalism.
Thus the topological algebra may be viewed as capturing the `invariant'
meaning of the theory, be it in the DDK formulation or in the guise of
constrained integrable \hs.

Let us note that, according to the formula \req{d(c)},
the minimal-model values of the matter central charge $d$
$$ d=1-{6(p\pr-p)^2\over p\pr p}$$
come from the rational values
$${\ctop\over3}=1-2\L({p\over p\pr}\R)^{\pm1}$$
of the \tcc,
so that for $p\pr$ or $p=1$, the corresponding formula
${\ctop\over3}=1-{2\over p}$
becomes
that for the $N=2$ superconformal central charge $c={k\over
k+2}$, $k=p-2$ (this point of view was elaborated recently in
\cite{[BLNW]}). The thus distinguished r\^ole of the
$(p,1)$ models may explain the fact that only for the $(l,1)$
decoupling operators there is no obstruction to the factorization
through the Virasoro constraints.

\bigskip

\noi
{\sc Acknowledgements.} We are grateful to L.~Alvarez-Gaum\'e,
B.~A.~Dubrovin, G.~Fel\-der, J.~Fr\"oh\-lich,
A.~Ga\-nchev, K.~Gaw\c{e}dzki, A.~Gorsky,
W.~Ler\-che, A.~Lossev,
A.~N.~Schel\-lek\-ens, A.~Turbiner and
M.~A.~Va\-si\-liev for useful
discussions. We are specially indebted
to R.~Dijkgraaf, E.~Kiritsis and A.~A.~Tseytlin
for very useful suggestions.
A.~S. wishes to thank E.~Corrigan for his kind
hospitality at the I.~Newton Institute, Cambridge.

\end{document}